\documentclass[twocolumn,floatfix,aps, prb]{revtex4-2}

\usepackage[utf8]{inputenc}
\usepackage{graphicx, bm,bbold}

\usepackage{amsmath}
\usepackage{amssymb}
\usepackage{color}
\usepackage{comment}

\usepackage{mathtools}
\usepackage[FIGTOPCAP]{subfigure}
\usepackage[]{footmisc}
\usepackage{dsfont}
\usepackage{soul,xcolor}
\usepackage[pdfborder={0 0 0}]{hyperref}
\usepackage{lineno}

\usepackage[normalem]{ulem}

\newcommand{\vk}{{\bm{k}}}
\newcommand{\vq}{\bm{q}}
\newcommand{\rr}{\bm{r}}

\newcommand{\vs}{\boldsymbol{\sigma}}

\newcommand{\pd}{{\phantom{\dagger}}}
\newcommand{\en}{\varepsilon}

\begin{document}

\title{Axionic Instability of Periodic Weyl-Semimetal Superstructures}

\author{Tommy Li}
\author{Maxim Breitkreiz}
\email{breitkr@physik.fu-berlin.de}
\affiliation{Dahlem Center for Complex Quantum Systems and Fachbereich Physik, Freie Universit\" at Berlin, 14195 Berlin, Germany}

\date{April 2024}

\begin{abstract}
Weyl-semimetal superstructures with a spiraling position of a pair of Weyl nodes of opposite chirality can host a chiral-symmetry preserving Fermi-arc metal state, where the chirality is carried by cylindrical Fermi surfaces, electron- and hole-like  depending on the chirality. The Fermi surfaces nest at vanishing momentum separation (zero nesting vector) at the  electron-hole-compensation energy because the nesting 
is topologically protected by vanishing spatial overlap  of any pair of equal-momentum opposite-chirality states.
In this work we show that the nesting and Coulomb interaction  drive a spontaneous chiral symmetry 
breaking in such a Fermi arc metal, which leads 
 to a dynamical axion insulator state but without breaking 
 translational symmetry (no charge-density-wave order) as in a conventional Weyl semimetal. 
As for material realization, we discuss magnetically doped Bi$_2$Se$_3$, for which the Weyl-node positions depend on the order of the magnetic dopands. In this case, the
axionic condensation can itself stabilize a spiral order of the magnetization, and hence the spiraling node positions, even if the magnetic interaction is intrinsically ferromagnetic.
\end{abstract}

\maketitle

\emph{Introduction}---The axion is the Goldstone mode associated with 
spontaneous chiral symmetry breaking, which has been considered as a candidate for dark matter \cite{Preskill:1982cy,Abbott:1982af,Dine:1982ah} and as a solution of the strong CP problem
\cite{PhysRevLett.38.1440,Weinberg:1977ma,Wilczek:1977pj}.
Its condensed-matter realization, in form of a dynamical quasiparticle mode
$\theta$ coupled to the electromagnetic
field in the form $\theta\, \bm{E}\cdot\bm{B}$
\cite{Li2010}, is in great demand from the 
perspective of realizing novel topological states of matter. The dynamical condensed-matter axion is associated with novel electronic response properties \cite{Li2010,Wang2013,Bernevig2022}, 
as well as the intriguing possibility to build dark-matter detectors 
\cite{Marsh2019, Chigusa2021, Semertzidis2022, Schutte_Engel_2021}. 

Theoretical proposals to realize  axions in condensed matter include the realization of 
massive axions in antiferromagnetic topological insulators \cite{Li2010} and massless axions
in Weyl semimetals (WSMs) \cite{Wang2013, Roy2015}. The experimental realization is so far limited to the WSM (TaSe$_4$)$^2$I \cite{Colletta2013,Gooth2019,Shi2021}.
In WSMs the low-energy excitations resemble chiral Weyl fermions (chirality $\chi=\pm$) represented by the Hamiltonian
\begin{gather}
H_{\bm{k}\chi} = \chi \,v\,\vs\cdot \vk \ \ , 
\label{eq1}
\end{gather}
where $\vs$ is the vector of (pseudo-) spin Pauli matrices, 
$\vk$ the momentum ($\hbar=1$), and $v$ the Fermi velocity. The Weyl nodes occuring in pairs of opposite chirality at different momenta are topologically protected on length scales much larger than the inverse separation of different chiralities in momentum space. Excitonic- or Peierls-type instabilities \cite{Keldysh1964,Peierls1955,Jerome1967} leading to pairing of opposite chiralities constitute a spontaneous chiral symmetry breaking and thus lead to axions  \cite{Wang2013, Wei2012, Roy2015, You2016}. However, the momentum-space separation of opposite chiralities necessitates an accompanying breaking of translational symmetry manifesting in a charge-density-wave formation, the  phase of which (phason) is the massless axion field. The pure axion electrodynamics is obscured in these systems, since the pinning of the charge-density wave by impurities leads to a threshold voltage \cite{Gruner1988a,Gooth2019} below which the axion remains static. Moreover, the charge density wave couples to the electric field via additional terms beyond the characteristic axion coupling \cite{Gooth2019}. 

The extension of this so far experimentally most successful realization of axions to other WSM materials meets the difficulties that purely excitonic 
internodal pairing does not gap the Weyl nodes \cite{Wei2012} and, due to the 
vanishing density of states at the Weyl nodes, a minimal
interaction strength \cite{ Wei2012,Wang2013,Wei2014,Xue2017} or the application of a magnetic field  \cite{Yang2011, Roy2015} is required.

In this work we propose an alternative condensed-matter platform for realizing axions, which has a number of advantages as compared to WSMs, among which is that the spontanous chiral symmetry breaking is not accompanied by a charge density wave and does not require an external magnetic field. The host system is a Fermi-arc metal (FAM) -- a WSM superstructure with spatially modulated Weyl-node positions, recently predicted by Brouwer and one of the authors 
\cite{Breitkreiz2022}. FAMs are topological semimetallic states  where instead of Weyl nodes the Weyl-fermion chirality is carried by extended Fermi surfaces. Crucially, here the topological protection is based not on a purely momentum-like separation of chiralities as in the original WSM but a mixed momentum-real-space separation, explained in more detail below.
This momentum-real-space separation allows for protected surfaces of degenerate states  of opposite chirality and opposite electron-hole character in momentum space. We show that these states are unstable with respect to the formation of an excitonic electron-hole condensate that spontaneously breaks chiral symmetry. This leads to dynamic axions as the Goldstone modes associated with the chiral-symmetry breaking. Since the  paired electrons and holes are at the same momentum, translational symmetry remains intact so that no charge density wave is formed. This makes FAMs a more suitable platform to realize pure axion electrodynamics than WSMs. The instability that we will explore in this work is of the type of a nesting-driven excitonic instability  \cite{Keldysh1964, Kulikov1984, Fawcett1988a}, but modified in accordance with the  fermiology of the FAMs. Next, we will present our considerations using one specific FAM model and we will discuss generalizations to other FAM systems at the end.

\begin{figure}
    \centering
    \includegraphics[width=0.7\columnwidth]{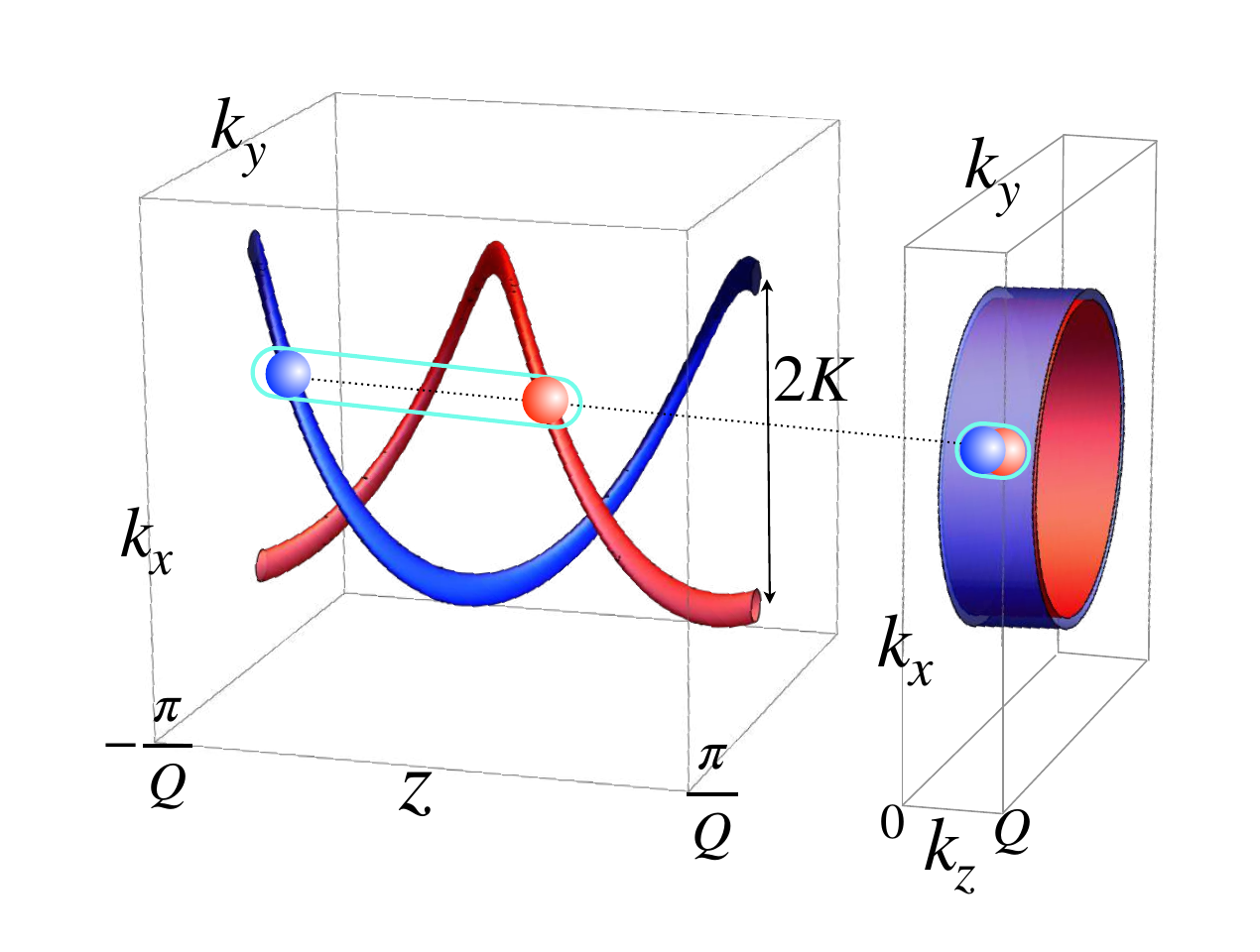}
    \caption{Left: Fermi-arc metal states at fixed low energy in mixed momentum-real space $(k_x,k_y,z)$ shown as position of the wavefunction centers $z_{\bm{k}n\chi}$ (red/blue helical 
    lines for chirality $\chi=+/-$). Right: Same states but in the all-momentum space $(k_x,k_y,k_z)$, where they form nested cylindrical electron- and hole-like Fermi surfaces for chirality $+$ and $-$, respectively. Both figures illustrate excitonic pairing of two states (red/blue speheres) of opposite chirality at the Fermi surface, whih are at the same momentum (right)
    but separated in $z$ by $\pi/Q$ (left), which protects them from single-particle hybridization.}
    \label{fig1}
\end{figure}

\emph{Model}---The model system we consider is based on the model of a magnetically doped 3D topological insulator of the Bi$_2$Se$_3$ family \cite{Cho2011,Liu2013b,Vazifeh2013}, in the ferromagnetic Weyl-semimetal phase (exchange coupling  larger than the band gap of the paramagnetic phase).
 The low-energy 
Hamiltonian is given by \eqref{eq1} with added magnetic 
exchange term that splits the two Weyl nodes of opposite chirality to $\bm{k}=\pm \bm{K}$,
\begin{gather}
    H = \chi\, v\,\bm{\sigma}\cdot \bm{k} + v\,\bm{\sigma}\cdot\bm{K}.
\end{gather}
The FAM phase emerges if the ferromagnetic order of the magnetic dopands is 
replaced by a helical magnetic order, in which the exchange field 
rotates as a function of one spatial coordinate 
(here $z$) perpendicular to the rotation plane (here $\vk_\parallel =(k_x,k_y)$) in momentum space, 
\begin{gather}
\bm{K} \to \bm{K}(z) = K(\cos(Qz), \sin(Qz),0),
\end{gather}
where the pitch $2\pi/Q$ is assumed sufficiently long so that $Q\ll K\equiv |\bm{K}|$.
Such a spiral magnetic order can arise from 
Weyl-fermion mediated magnetic interaction \cite{Chang2015a,Wang2017a,Nikolic2021,Nikolic2021b};
recently, WSMs with spiral magnetic order have been observed
experimentally \cite{Gaudet2021,Yao2022,Drucker2023}. Moreover, the excitonic condensation considered in this work can by itself stabilize the spiral magnetic order against the ferromagnetic, which will be discussed in more detail in section ``Free-energy contribution of axionix condensate''. 

To understand the fermiology of FAMs (see Ref.\ \cite{Breitkreiz2022} for more details) it is useful to assume for a moment that $z$ in $\bm{K}(z)$ is a parameter, in which case the system is a WSM where the Weyl-node positions spiral in form of a double helix in the mixed ($\vk_\parallel,z$) momentum/real space, as illustrated in Fig.\ \ref{fig1} (left). Low-energy states can only live nearby those Weyl node positions. At the same time, the spatial node-position change induces a pseudo-magnetic field $\nabla\times\bm{K}(z)$ since $\bm{K}(z)$ enters the Hamiltonian as a pseudo vector potential (with opposite signs for opposite chiralities). The states become localized in $z$ and the pure Weyl-Fermion dispersion is replaced by pseudo Landau levels dispersing along $\bm{K}(z)$. This is reflected in the exact low-energy dispersion of the FAM \cite{Breitkreiz2022},
\begin{gather}
\varepsilon_{\bm{k}\chi} = \chi\, v(k_\parallel-K) \ \ , \end{gather}
where we used $\vk_\parallel=k_\parallel (\cos\phi_\vk,\sin\phi_\vk)$. The corresponding eigenstates
read
\begin{gather}
\psi_{\bm{k}\chi} = \bigg(\frac{\pi K}{4Q}\bigg)^{\frac{1}{4}}\left(\begin{array}{c} 1\\ - e^{i\phi_{\bm{k}}}\end{array}\right)\times \nonumber\\
\sum_n e^{i(\bm{k}_\parallel \cdot\bm{r}_\parallel +k_z z_{\bm{k}n\chi})} e^{-\frac{KQ}{2}(z - z_{\bm{k},n,\chi})^2} \ \ , \nonumber\\
z_{\bm{k}n+} =
\frac{\phi_{\bm{k}} + 2\pi n}{Q} \ \ ,  \ \  z_{\bm{k}n-} = z_{\bm{k}n+} + \frac{\pi}{Q} \ \ ,
\label{wf}
\end{gather}
where $n$ indexes the unit cells of size $2\pi/Q$ of the helix periodicity in the $z$ direction.
The two chiralities $\chi=\pm$ correspond to electron- and hole-like cylindrical Fermi surfaces with radius $K$ in the $k_x,k_y$ plane and flat in $k_z$, shown in Fig.\ \ref{fig1} (right). At the same time, due to localization of the wavefunctions in the
$z$ direction, the quantum number $k_z$ maybe replaced by the unit-cell index $n$, resulting in an equivalent representation of the Fermi surface in the mixed momentum-real space $(k_x,k_y,z)$, Fig.\ \ref{fig1} (left), where the Fermi surfaces form  a double helix.
Thus, at electron-hole compensation energy (here $\varepsilon=0$) the two opposite-chirality Fermi surfaces nest without any momentum
separation. In the mixed ($\vk_\parallel,z$) space, however, the chiralities are instead always separated up to an exponentially small overlap $\sim e^{-K/Q}$ ($Q\ll K$), which makes the protection of the nesting without momentum separation
 from single-particle hybridization. 

This protected overlap of chiralities in the energy-momentum space 
distinguishes the FAMs from the
 WSMs, where the topological protection of chiralities relies on separation of opposite chiralities in momentum ($\sim K$).  
Crucially, while two quasiparticles of opposite chirality at the same momentum are spatially separated and thus don't hybridize on the single-particle level, there is still Coulomb interaction between them. Moreover, the nesting of the electron- and hole-like Fermi surfaces, characterized by the corresponding dispersions fulfilling 
$\varepsilon_{+,\vk} =-\varepsilon_{-,\vk+\bm{{\cal Q}}}$ is well-known to support excitonic-type instabilities  \cite{Keldysh1964, Kulikov1984}. The nesting vector $\bm{{\cal Q}}$ is zero in our case, unlike in conventional excitonic systems and WSMs \cite{Wang2013}.

\emph{Excitonic instability}---We now show that the FAM exhibits an excitonic-type instability for arbitrarily weak values of the interaction caused by collisions between electrons and holes of opposite chirality. This instability is significantly modified by the spatial structure of the wavefunctions (\ref{wf}) which for $Q \ll K$ strongly suppresses scattering unless the in-plane momentum transfer is close to zero. Since the matrix elements $U_{\bm{q}} = 4\pi e^2/q^2$ of the Fourier-transformed long ranged Coulomb interaction diverge for small momentum transfer, the excitonic instability is also strongly affected by screening processes, which we incorporate within the random phase approximation (RPA). 

We may describe the mean-field instability via the excitonic self-energy $\Delta(\bm{k},\omega)$, which satisfies the self-consistency equation
\begin{gather}
\Delta_{+-}(\bm{k},\omega) = \sum_{\omega',\bm{k}'}\Gamma_{+-}(\bm{k},\bm{k}',\omega-\omega') G_{+-}(\bm{k}',\omega'),
\label{MF}
\end{gather}
where $G_{+-}(\bm{k},\omega) = - \int \langle \mathcal{T} c^\pd_{\bm{k},+}(t) c^\dag_{\bm{k},-}(0)\rangle e^{i\omega t} dt$ is the off-diagonal matrix element of the Matsubara Green's function in the chirality basis; $c^\dag_{\bm{k},\chi}$ ($c^\pd_{\bm{k},\chi}$  ) are the fermionic creation (annihilation) operators,  and $\Gamma_{+-}(\bm{k}',\Omega)$ is the scattering vertex. In order to remove the divergence of the forward scattering vertex due to the long-ranged nature of the Coulomb interaction, we consider screening in the RPA. Hence, the vertex is given by [detailed derivation in Appendix A and B],
\begin{gather}
\Gamma_{+-}(\bm{p},\bm{k},\Omega) =\frac{U_{\bm{q}}}{1 -\Pi(\bm{q},\Omega)U_{\bm{q}}} f(\bm{p},\bm{k}),
\end{gather}
where $\bm{q}= \bm{p}-\bm{k}$, $\Pi(\bm{q},\Omega)$ is the polarization operator, and the form factor 
\begin{gather}
  f(\bm{p},\bm{k})=   \frac{1 + e^{i(\phi_{\bm{p}} - \phi_{\bm{k}})}}{2} e^{-\frac{K}{2Q} (\phi_{\bm{p}} - \phi_{\bm{k}})^2}
\end{gather}
stems from the overlap of wavefunctions given in \eqref{wf}. The physical reason for the unusual form factor is that two wavefunctions at different momenta $\vk$ and $\vk'$  with $|\phi_\vk -\phi_\vk'|\gtrsim \sqrt{Q/K}$ are localized at different $z$ slices so that they don't interfere. Hence, density waves with wavevectors $\vk-\vk'$ for which $|\phi_\vk -\phi_\vk'|\gtrsim \sqrt{Q/K}$ become suppressed.

Approaching the critical temperature for excitonic condensation, $T\rightarrow T_c$, we may replace $G_{+-}(\bm{k}',\omega') \rightarrow G_+(\bm{k}',\omega')\Delta_{+-}(\bm{k}',\omega') G_-(\bm{k}',\omega')$, with normal state single electron Green's functions $G_\pm(\bm{k}',\omega') = (i\omega'-\varepsilon_{\bm{k}',\pm})^{-1}$. Due to the nesting of the electron and hole Fermi surfaces, contributions to the frequency and momentum summation in (\ref{MF}) originating from the singularities of the normal state single electron Green's functions $G_\pm(\bm{k}',\omega')$ for $i\omega' = \varepsilon_{\bm{k}',\chi}\approx 0$ become enhanced by factors $\log(w/T)$ with $w$ being an ultraviolet energy cut-off. Physically, this corresponds to an electron-hole instability due to interaction processes close to the Fermi surface, which we may treat similarly to the BCS theory of superconductivity \cite{Bardeen1957}. We neglect subleading corrections that do not affect the logarithmic enhancement, corresponding to virtual processes that involve electron-hole pairs with energy far above or below the Fermi level, which allows us to set $\omega, \Omega = 0$ in the excitonic self-energy and interaction vertex respectively.

The static polarization operator is given by
\begin{gather}
\Pi(\bm{q},\Omega=0) = \sum_{\bm{k},\chi}
\frac{n_F(\varepsilon_{\bm{k},\chi}) - n_F(\varepsilon_{\bm{k}+\bm{q},\chi})}{\varepsilon_{\bm{k},\chi}- \varepsilon_{\bm{k}+\bm{q},\chi}} f(\bm{k}+\bm{q},\bm{k}) \ \ ,
\end{gather}
where $n_F(x)=1/(1+\exp{(x/T)})$ is the Fermi-Dirac distribution and the Boltzmann constant set to unity. It describes the density perturbation induced by a static external potential. Excitation of electron-hole pairs is suppressed unless the total in-plane momentum is small, and we find that it decreases as $\Pi(\bm{q},\Omega=0) \sim (KQ)^{\frac{3}{2}}/v |\bm{q}_\parallel|$, while in the limit $|\bm{q}_\parallel| \ll \sqrt{KQ}$, we recover the Thomas-Fermi limit
\begin{gather}
\Pi(\bm{q},\Omega=0)\rightarrow -\nu = -\frac{KQ}{2\pi^2 v} \ \ ,
\end{gather}
where $\nu=KQ/2\pi^2v$ is the density of states at the Fermi energy.
Since the summation appearing in the self-consistency equation (\ref{MF}) is dominated by contributions where $\bm{k}'_\parallel \approx \bm{k}_\parallel$, we may replace the scattering vertex with the simplified expression
\begin{gather}
\Gamma_{+-}(\bm{k},\bm{k}') \rightarrow \frac{U_{\bm{k}-\bm{k}'}}{1 + \nu U_{\bm{k}-\bm{k}'}}f(\bm{k}',\bm{k})\ \ .
\end{gather}

The critical temperature is determined from the gap equation \eqref{MF} following standard steps \cite{Bardeen1957} (detailed in Appendix C),
\begin{gather}
T_c = \frac{2e^\gamma}{\pi} w \exp \left[-\frac{1}{\eta}\sqrt{\frac{K}{Q}}\right] \ \ ,
\nonumber\\
\eta = \frac{\xi e^{\xi^2}\text{erf}(\xi)}{2\sqrt{2}} \ \ , \ \ \ 
\xi^2 =  \frac{\alpha\, c}{\pi\, v}, \ \ 
\end{gather}
where $\gamma$ is the Euler constant,  $c$ is the speed of light, and  $\alpha$ is the fine-structure constant.  Analogous to other excitonic systems with finite density of states \cite{Jerome1967} and the BCS theory \cite{Bardeen1957} no minimal interaction strength is required, in contrast to   WSMs, where a minimal interaction strength emerges due to the vanishing density of states at the Weyl nodes \cite{Wei2012,Wang2013,Wei2014,Xue2017}. 

Below $T_c$ the mean field $\Delta_{+-}$ couples opposite chiralities as expressed by \eqref{MF}. This spontaneous chiral symmetry breaking leads to axions analogous to conventional WSMs (but with nesting vector set to zero). In a nutshell (see, e.g., Ref.\ \cite{Wang2013} for details), the chiral gauge transformation can be used to remove the phase of $\Delta_{+-}$ at the cost of adding an anomalous term in the action from the path-integral measure, which is of the form $\sim\theta\, \bm{E}\cdot\bm{B}$ \cite{Fujikawa1979,Wang2013,Schwartz2014}. Thus, the Goldstone mode of the excitonic condensate couples to the electro-magnetic field as an axion, detectable via specific axionic response properties \cite{Li2010,Wang2013,Bernevig2022}.

\emph{Free-energy contribution of axionic condensate}---In our model of magnetically doped topological insulator, the FAM state arises from a ferromagnetic WSM state if the ferromagnetic alignment of magnetic moments receives a smooth modulation towards a spiraling magnetic order.  
Let us now assume that if disregarding the excitonic condensation, the dominant interaction of the magnetic moments is ferromagnetic. Then, depending on the stiffness of the magnetic moments, the excitonic state of the FAM may  
have lower energy than the WSM state, in which case the spiral magnetic order will itself be driven by the excitonic instability.  

The free-energy contribution  of the excitonic condensate in the FAM state to fourth order in the order parameter $\Delta$ reads 
$F =\alpha|\Delta|^2 + (\beta/2) |\Delta|^4 $. The coefficients  below $T_c$ can be derived in the standard way 
(see Appendix D for a detailed derivation), $\alpha= \nu (T-T_c)/T$ and
$\beta= (7\zeta(3)/8\pi^2)\nu/T^2$.
Below $T_c$, the order parameter becomes $|\Delta|^2 =-\alpha/\beta $  and the free energy contribution of the condensate thus reads
\begin{gather}
    F_\Delta = -\frac{\alpha^2}{2\beta} = -\nu\big(T-T_c\big)^2\frac{4\pi^2}{7\zeta(3)}.
\end{gather}
Note that $F_\Delta$ depends on the magnitude of the magnetic order through the density of states $\nu=KQ/2\pi^2v$. 
This condensate contribution is to be compared to a positive $Q^2$ free energy contribution due to magnetic stiffness that would stabilize a ferromagnetic order (a ferromagnetic order would correspond to $Q=0$). Such a contribution can be written as $F_M =  (w^2/v)\rho_M Q^2$, where $\rho_M>0$ is a dimensionless paremeter 
quantifying the stiffness of the ferromagnetic order. 
The condition for the formation of the axionic insulator via the FAM is then a negative free energy at finite $Q$, which is achieved for 
\begin{equation}
 \rho_M  < \frac{8\,e^{2\gamma}}{7\zeta(3)} \frac{K}{Q} \,\bigg(1-\frac{T}{T_c}\bigg)^2\, \exp\bigg(-\frac{2}{\eta}\sqrt{\frac{K}{Q}}\bigg),
\end{equation}
defining an upper bound for the stiffness.

\emph{Discussion and conclusion}---We have shown that periodic WSM superstructures in a FAM state that support nested Fermi surfaces are unstable with respect to excitonic condensation spontaneously breaking chiral symmetry. These systems are suitable to realize dynamic axions as the Goldstone modes of the symmetry breaking. Compared to WSMs, which are the only systems where dynamical axions have been experimentally realized so far, the realization of axions in FAMs has the advantages that neither a minimal interaction strength \cite{Wang2013} nor an external magnetic field \cite{Roy2015} are required and, most importantly, the axion dynamics is not accompanied by the dynamics of a charge density wave  \cite{Colletta2013,Gooth2019,Shi2021}.   These advantages are the result of the novel form in which the chiral anomaly is realized in FAMs as compared to WSMs: The density of states is finite at the electron-hole compensation point (the Fermi surfaces don't shrink to points as in WSMs) and the topological protection of chiralities is based not on a purely momentum-like separation of chiralities as in WSMs but a mixed momentum-real-space separation, which allows for nested Fermi surfaces of opposite chirality at the same momentum. 

We have explicitely considered the model of a magnetically doped 3D topological insulator from the Bi$_2$Se$_3$ family, which has been proposed as a WSM provided a ferromagnetic magnetic order \cite{Cho2011,Liu2013b,Vazifeh2013}. The FAM is formed in a similar system but with a spiral magnetic order instead of a ferromagnetic one \cite{Breitkreiz2022}. The spiral magnetic order can be stabilized by mechanisms  considered elsewhere \cite{Chang2015a,Wang2017a,Nikolic2021,Nikolic2021b} but it can also be stabilized by the excitonic instability itself, as we have shown in this work. We derived the uppper bound for the magnetic stiffness of the ferromagnetic coupling, 
$\sim (K/Q) \exp\big(-\sqrt{K/Q}\big)$, up to which the excitonic instability would lead to the spiral magnetic order, according to our mean-field considerations. Here one should note that we assumed clean systems with $Q\ll K$ being the only requirement on $Q$. In real systems there will be also a lower bound for $Q$ due to disorder, as the  mean free path $l$ naturally limits the size of the unit cell in the long $z$ direction, $2\pi/Q$, up to which disorder can be treated as a perturbation to the FAM state, hence $Q\gg l^{-1}$ is an additional condition on $Q$.

Material realizations other than magnetically doped Bi$_2$Se$_3$ can be discussed based on 
the generic ingredients favoring the FAM state, such as  the coupling of the magnetization direction and the momentum separation of Weyl nodes \cite{Weststrom2017,Ilan2020,Araki2020} and coupling mechanisms of the magnetic moments that favor a spiral magnetic order. Interestingly, the latter can be provided by Weyl-mediated magnetic interactions as predicted theoretically 
\cite{Chang2015a,Wang2017a,Nikolic2021,Nikolic2021b, Duan2018, Kaladzhyan2019a, Verma2020}
and observed experimentally \cite{Gaudet2021,Yao2022,Drucker2023}. 
In particular, suitable material platforms include
van der Waals layered magnets from the MnBi$_2$Te$_4$-family  compounds \cite{A02Li2019,Otrokov2019MBT}, showing a strongly manipulable magnetic coupling, and magnetically doped Dirac semimitals \cite{Deng2017,Cano2017}. Here too,  a slow spiraling modulation of magnetic order can lead to FAMs.

 In these different material realizations of FAMs, the Fermi surfaces may differ from the cylindrical shape of the model considered in this work.  While a precise prediction for such systems requires modified FAM models, we note that the crucial property of FAMs to allow hybiridization-protected states of opposite chirality overlapping in energy-momentum is generic and thus, in principle, possible in form of arbitrary Fermi-surface shapes. The system symmetry needs to support nesting though, ideally at a vanishing momentum separation to not link the chiral symmetry breaking with translational symmetry breaking. The cylindrical shape of the Fermi surfaces of our model are instead not essential, since the excitonic pairing is limited to contributions separated no
larger than $\sqrt{KQ}$ which is much smaller than the cylinder radius $K$.

Intriguingly, further theoretical exploration might reveal correlation effects that compete with or supplement the  excitonic instability considered in this work, such as
Peierls-type instabilities building on the here-neglected coupling of quasiparticles with phonons, or axion-fluctuation driven superconducting pairing.

\emph{Acknowledgements}---We thank Jan Sch\" utte-Engel, Binghai Yan, and Carsten Timm for useful discussions. This research was funded by the Deutsche Forschungsgemeinschaft (DFG, German Research Foundation) through CRC-TR 183 “Entangled States of Matter” 
and the Emmy Noether program, Project No. 506208038.

\bibliography{library}	

\begin{thebibliography}{52}%
\makeatletter
\providecommand \@ifxundefined [1]{%
 \@ifx{#1\undefined}
}%
\providecommand \@ifnum [1]{%
 \ifnum #1\expandafter \@firstoftwo
 \else \expandafter \@secondoftwo
 \fi
}%
\providecommand \@ifx [1]{%
 \ifx #1\expandafter \@firstoftwo
 \else \expandafter \@secondoftwo
 \fi
}%
\providecommand \natexlab [1]{#1}%
\providecommand \enquote  [1]{``#1''}%
\providecommand \bibnamefont  [1]{#1}%
\providecommand \bibfnamefont [1]{#1}%
\providecommand \citenamefont [1]{#1}%
\providecommand \href@noop [0]{\@secondoftwo}%
\providecommand \href [0]{\begingroup \@sanitize@url \@href}%
\providecommand \@href[1]{\@@startlink{#1}\@@href}%
\providecommand \@@href[1]{\endgroup#1\@@endlink}%
\providecommand \@sanitize@url [0]{\catcode `\\12\catcode `\$12\catcode
  `\&12\catcode `\#12\catcode `\^12\catcode `\_12\catcode `\%12\relax}%
\providecommand \@@startlink[1]{}%
\providecommand \@@endlink[0]{}%
\providecommand \url  [0]{\begingroup\@sanitize@url \@url }%
\providecommand \@url [1]{\endgroup\@href {#1}{\urlprefix }}%
\providecommand \urlprefix  [0]{URL }%
\providecommand \Eprint [0]{\href }%
\providecommand \doibase [0]{https://doi.org/}%
\providecommand \selectlanguage [0]{\@gobble}%
\providecommand \bibinfo  [0]{\@secondoftwo}%
\providecommand \bibfield  [0]{\@secondoftwo}%
\providecommand \translation [1]{[#1]}%
\providecommand \BibitemOpen [0]{}%
\providecommand \bibitemStop [0]{}%
\providecommand \bibitemNoStop [0]{.\EOS\space}%
\providecommand \EOS [0]{\spacefactor3000\relax}%
\providecommand \BibitemShut  [1]{\csname bibitem#1\endcsname}%
\let\auto@bib@innerbib\@empty
\bibitem [{\citenamefont {Preskill}\ \emph {et~al.}(1983)\citenamefont
  {Preskill}, \citenamefont {Wise},\ and\ \citenamefont
  {Wilczek}}]{Preskill:1982cy}%
  \BibitemOpen
  \bibfield  {author} {\bibinfo {author} {\bibfnamefont {J.}~\bibnamefont
  {Preskill}}, \bibinfo {author} {\bibfnamefont {M.~B.}\ \bibnamefont {Wise}},\
  and\ \bibinfo {author} {\bibfnamefont {F.}~\bibnamefont {Wilczek}},\
  }\bibfield  {title} {\bibinfo {title} {{Cosmology of the Invisible Axion}},\
  }\href {https://doi.org/10.1016/0370-2693(83)90637-8} {\bibfield  {journal}
  {\bibinfo  {journal} {Phys. Lett. B}\ }\textbf {\bibinfo {volume} {120}},\
  \bibinfo {pages} {127} (\bibinfo {year} {1983})}\BibitemShut {NoStop}%
\bibitem [{\citenamefont {Abbott}\ and\ \citenamefont
  {Sikivie}(1983)}]{Abbott:1982af}%
  \BibitemOpen
  \bibfield  {author} {\bibinfo {author} {\bibfnamefont {L.~F.}\ \bibnamefont
  {Abbott}}\ and\ \bibinfo {author} {\bibfnamefont {P.}~\bibnamefont
  {Sikivie}},\ }\bibfield  {title} {\bibinfo {title} {{A Cosmological Bound on
  the Invisible Axion}},\ }\href {https://doi.org/10.1016/0370-2693(83)90638-X}
  {\bibfield  {journal} {\bibinfo  {journal} {Phys. Lett. B}\ }\textbf
  {\bibinfo {volume} {120}},\ \bibinfo {pages} {133} (\bibinfo {year}
  {1983})}\BibitemShut {NoStop}%
\bibitem [{\citenamefont {Dine}\ and\ \citenamefont
  {Fischler}(1983)}]{Dine:1982ah}%
  \BibitemOpen
  \bibfield  {author} {\bibinfo {author} {\bibfnamefont {M.}~\bibnamefont
  {Dine}}\ and\ \bibinfo {author} {\bibfnamefont {W.}~\bibnamefont
  {Fischler}},\ }\bibfield  {title} {\bibinfo {title} {{The Not So Harmless
  Axion}},\ }\href {https://doi.org/10.1016/0370-2693(83)90639-1} {\bibfield
  {journal} {\bibinfo  {journal} {Phys. Lett. B}\ }\textbf {\bibinfo {volume}
  {120}},\ \bibinfo {pages} {137} (\bibinfo {year} {1983})}\BibitemShut
  {NoStop}%
\bibitem [{\citenamefont {Peccei}\ and\ \citenamefont
  {Quinn}(1977)}]{PhysRevLett.38.1440}%
  \BibitemOpen
  \bibfield  {author} {\bibinfo {author} {\bibfnamefont {R.~D.}\ \bibnamefont
  {Peccei}}\ and\ \bibinfo {author} {\bibfnamefont {H.~R.}\ \bibnamefont
  {Quinn}},\ }\bibfield  {title} {\bibinfo {title} {{CP Conservation in the
  Presence of Pseudoparticles}},\ }\href
  {https://doi.org/10.1103/PhysRevLett.38.1440} {\bibfield  {journal} {\bibinfo
   {journal} {Phys. Rev. Lett.}\ }\textbf {\bibinfo {volume} {38}},\ \bibinfo
  {pages} {1440} (\bibinfo {year} {1977})}\BibitemShut {NoStop}%
\bibitem [{\citenamefont {Weinberg}(1978)}]{Weinberg:1977ma}%
  \BibitemOpen
  \bibfield  {author} {\bibinfo {author} {\bibfnamefont {S.}~\bibnamefont
  {Weinberg}},\ }\bibfield  {title} {\bibinfo {title} {{A New Light Boson?}},\
  }\href {https://doi.org/10.1103/PhysRevLett.40.223} {\bibfield  {journal}
  {\bibinfo  {journal} {Phys. Rev. Lett.}\ }\textbf {\bibinfo {volume} {40}},\
  \bibinfo {pages} {223} (\bibinfo {year} {1978})}\BibitemShut {NoStop}%
\bibitem [{\citenamefont {Wilczek}(1978)}]{Wilczek:1977pj}%
  \BibitemOpen
  \bibfield  {author} {\bibinfo {author} {\bibfnamefont {F.}~\bibnamefont
  {Wilczek}},\ }\bibfield  {title} {\bibinfo {title} {{Problem of Strong $P$
  and $T$ Invariance in the Presence of Instantons}},\ }\href
  {https://doi.org/10.1103/PhysRevLett.40.279} {\bibfield  {journal} {\bibinfo
  {journal} {Phys. Rev. Lett.}\ }\textbf {\bibinfo {volume} {40}},\ \bibinfo
  {pages} {279} (\bibinfo {year} {1978})}\BibitemShut {NoStop}%
\bibitem [{\citenamefont {Li}\ \emph {et~al.}(2010)\citenamefont {Li},
  \citenamefont {Wang}, \citenamefont {Qi},\ and\ \citenamefont
  {Zhang}}]{Li2010}%
  \BibitemOpen
  \bibfield  {author} {\bibinfo {author} {\bibfnamefont {R.}~\bibnamefont
  {Li}}, \bibinfo {author} {\bibfnamefont {J.}~\bibnamefont {Wang}}, \bibinfo
  {author} {\bibfnamefont {X.~L.}\ \bibnamefont {Qi}},\ and\ \bibinfo {author}
  {\bibfnamefont {S.~C.}\ \bibnamefont {Zhang}},\ }\bibfield  {title} {\bibinfo
  {title} {{Dynamical axion field in topological magnetic insulators}},\ }\href
  {https://doi.org/10.1038/nphys1534} {\bibfield  {journal} {\bibinfo
  {journal} {Nat. Phys.}\ }\textbf {\bibinfo {volume} {6}},\ \bibinfo {pages}
  {284} (\bibinfo {year} {2010})}\BibitemShut {NoStop}%
\bibitem [{\citenamefont {Wang}\ and\ \citenamefont {Zhang}(2013)}]{Wang2013}%
  \BibitemOpen
  \bibfield  {author} {\bibinfo {author} {\bibfnamefont {Z.}~\bibnamefont
  {Wang}}\ and\ \bibinfo {author} {\bibfnamefont {S.~C.}\ \bibnamefont
  {Zhang}},\ }\bibfield  {title} {\bibinfo {title} {{Chiral anomaly, charge
  density waves, and axion strings from Weyl semimetals}},\ }\href
  {https://doi.org/10.1103/PhysRevB.87.161107} {\bibfield  {journal} {\bibinfo
  {journal} {Phys. Rev. B}\ }\textbf {\bibinfo {volume} {87}},\ \bibinfo
  {pages} {161107(R)} (\bibinfo {year} {2013})}\BibitemShut {NoStop}%
\bibitem [{\citenamefont {Bernevig}\ \emph {et~al.}(2022)\citenamefont
  {Bernevig}, \citenamefont {Felser},\ and\ \citenamefont
  {Beidenkopf}}]{Bernevig2022}%
  \BibitemOpen
  \bibfield  {author} {\bibinfo {author} {\bibfnamefont {B.~A.}\ \bibnamefont
  {Bernevig}}, \bibinfo {author} {\bibfnamefont {C.}~\bibnamefont {Felser}},\
  and\ \bibinfo {author} {\bibfnamefont {H.}~\bibnamefont {Beidenkopf}},\
  }\bibfield  {title} {\bibinfo {title} {{Progress and prospects in magnetic
  topological materials}},\ }\href {https://doi.org/10.1038/s41586-021-04105-x}
  {\bibfield  {journal} {\bibinfo  {journal} {Nature}\ }\textbf {\bibinfo
  {volume} {603}},\ \bibinfo {pages} {41} (\bibinfo {year} {2022})}\BibitemShut
  {NoStop}%
\bibitem [{\citenamefont {Marsh}\ \emph {et~al.}(2019)\citenamefont {Marsh},
  \citenamefont {Fong}, \citenamefont {Lentz}, \citenamefont {{\v{S}}mejkal},\
  and\ \citenamefont {Ali}}]{Marsh2019}%
  \BibitemOpen
  \bibfield  {author} {\bibinfo {author} {\bibfnamefont {D.~J.}\ \bibnamefont
  {Marsh}}, \bibinfo {author} {\bibfnamefont {K.~C.}\ \bibnamefont {Fong}},
  \bibinfo {author} {\bibfnamefont {E.~W.}\ \bibnamefont {Lentz}}, \bibinfo
  {author} {\bibfnamefont {L.}~\bibnamefont {{\v{S}}mejkal}},\ and\ \bibinfo
  {author} {\bibfnamefont {M.~N.}\ \bibnamefont {Ali}},\ }\bibfield  {title}
  {\bibinfo {title} {{Proposal to Detect Dark Matter using Axionic Topological
  Antiferromagnets}},\ }\href {https://doi.org/10.1103/PhysRevLett.123.121601}
  {\bibfield  {journal} {\bibinfo  {journal} {Phys. Rev. Lett.}\ }\textbf
  {\bibinfo {volume} {123}},\ \bibinfo {pages} {121601} (\bibinfo {year}
  {2019})}\BibitemShut {NoStop}%
\bibitem [{\citenamefont {Chigusa}\ \emph {et~al.}(2021)\citenamefont
  {Chigusa}, \citenamefont {Moroi},\ and\ \citenamefont
  {Nakayama}}]{Chigusa2021}%
  \BibitemOpen
  \bibfield  {author} {\bibinfo {author} {\bibfnamefont {S.}~\bibnamefont
  {Chigusa}}, \bibinfo {author} {\bibfnamefont {T.}~\bibnamefont {Moroi}},\
  and\ \bibinfo {author} {\bibfnamefont {K.}~\bibnamefont {Nakayama}},\
  }\bibfield  {title} {\bibinfo {title} {{Axion/hidden-photon dark matter
  conversion into condensed matter axion}},\ }\href
  {https://doi.org/10.1007/JHEP08(2021)074} {\bibfield  {journal} {\bibinfo
  {journal} {J. High Energy Phys.}\ }\textbf {\bibinfo {volume} {2021}}\bibinfo
   {number} { (8)},\ \bibinfo {pages} {74}}\BibitemShut {NoStop}%
\bibitem [{\citenamefont {Semertzidis}\ and\ \citenamefont
  {Youn}(2022)}]{Semertzidis2022}%
  \BibitemOpen
\bibfield  {number} {  }\bibfield  {author} {\bibinfo {author} {\bibfnamefont
  {Y.~K.}\ \bibnamefont {Semertzidis}}\ and\ \bibinfo {author} {\bibfnamefont
  {S.~W.}\ \bibnamefont {Youn}},\ }\bibfield  {title} {\bibinfo {title} {{Axion
  dark matter: How to see it?}},\ }\href
  {https://doi.org/10.1126/sciadv.abm9928} {\bibfield  {journal} {\bibinfo
  {journal} {Sci. Adv.}\ }\textbf {\bibinfo {volume} {8}},\ \bibinfo {pages}
  {eabm9928} (\bibinfo {year} {2022})}\BibitemShut {NoStop}%
\bibitem [{\citenamefont {Sch{\"{u}}tte-Engel}\ \emph
  {et~al.}(2021)\citenamefont {Sch{\"{u}}tte-Engel}, \citenamefont {Marsh},
  \citenamefont {Millar}, \citenamefont {Sekine}, \citenamefont {Chadha-Day},
  \citenamefont {Hoof}, \citenamefont {Ali}, \citenamefont {Fong},
  \citenamefont {Hardy},\ and\ \citenamefont
  {{\v{S}}mejkal}}]{Schutte_Engel_2021}%
  \BibitemOpen
  \bibfield  {author} {\bibinfo {author} {\bibfnamefont {J.}~\bibnamefont
  {Sch{\"{u}}tte-Engel}}, \bibinfo {author} {\bibfnamefont {D.~J.~E.}\
  \bibnamefont {Marsh}}, \bibinfo {author} {\bibfnamefont {A.~J.}\ \bibnamefont
  {Millar}}, \bibinfo {author} {\bibfnamefont {A.}~\bibnamefont {Sekine}},
  \bibinfo {author} {\bibfnamefont {F.}~\bibnamefont {Chadha-Day}}, \bibinfo
  {author} {\bibfnamefont {S.}~\bibnamefont {Hoof}}, \bibinfo {author}
  {\bibfnamefont {M.~N.}\ \bibnamefont {Ali}}, \bibinfo {author} {\bibfnamefont
  {K.~C.}\ \bibnamefont {Fong}}, \bibinfo {author} {\bibfnamefont
  {E.}~\bibnamefont {Hardy}},\ and\ \bibinfo {author} {\bibfnamefont
  {L.}~\bibnamefont {{\v{S}}mejkal}},\ }\bibfield  {title} {\bibinfo {title}
  {{Axion quasiparticles for axion dark matter detection}},\ }\href
  {https://doi.org/10.1088/1475-7516/2021/08/066} {\bibfield  {journal}
  {\bibinfo  {journal} {J. Cosmol. Astropart. Phys.}\ }\textbf {\bibinfo
  {volume} {2021}}\bibinfo  {number} { (08)},\ \bibinfo {pages}
  {66}}\BibitemShut {NoStop}%
\bibitem [{\citenamefont {Roy}\ and\ \citenamefont {Sau}(2015)}]{Roy2015}%
  \BibitemOpen
\bibfield  {number} {  }\bibfield  {author} {\bibinfo {author} {\bibfnamefont
  {B.}~\bibnamefont {Roy}}\ and\ \bibinfo {author} {\bibfnamefont {J.~D.}\
  \bibnamefont {Sau}},\ }\bibfield  {title} {\bibinfo {title} {{Magnetic
  catalysis and axionic charge density wave in Weyl semimetals}},\ }\href
  {https://doi.org/10.1103/PhysRevB.92.125141} {\bibfield  {journal} {\bibinfo
  {journal} {Phys. Rev. B}\ }\textbf {\bibinfo {volume} {92}},\ \bibinfo
  {pages} {125141} (\bibinfo {year} {2015})}\BibitemShut {NoStop}%
\bibitem [{\citenamefont {Tournier-Colletta}\ \emph {et~al.}(2013)\citenamefont
  {Tournier-Colletta}, \citenamefont {Moreschini}, \citenamefont {Aut{\`{e}}s},
  \citenamefont {Moser}, \citenamefont {Crepaldi}, \citenamefont {Berger},
  \citenamefont {Walter}, \citenamefont {Kim}, \citenamefont {Bostwick},
  \citenamefont {Monceau}, \citenamefont {Rotenberg}, \citenamefont {Yazyev},\
  and\ \citenamefont {Grioni}}]{Colletta2013}%
  \BibitemOpen
  \bibfield  {author} {\bibinfo {author} {\bibfnamefont {C.}~\bibnamefont
  {Tournier-Colletta}}, \bibinfo {author} {\bibfnamefont {L.}~\bibnamefont
  {Moreschini}}, \bibinfo {author} {\bibfnamefont {G.}~\bibnamefont
  {Aut{\`{e}}s}}, \bibinfo {author} {\bibfnamefont {S.}~\bibnamefont {Moser}},
  \bibinfo {author} {\bibfnamefont {A.}~\bibnamefont {Crepaldi}}, \bibinfo
  {author} {\bibfnamefont {H.}~\bibnamefont {Berger}}, \bibinfo {author}
  {\bibfnamefont {A.~L.}\ \bibnamefont {Walter}}, \bibinfo {author}
  {\bibfnamefont {K.~S.}\ \bibnamefont {Kim}}, \bibinfo {author} {\bibfnamefont
  {A.}~\bibnamefont {Bostwick}}, \bibinfo {author} {\bibfnamefont
  {P.}~\bibnamefont {Monceau}}, \bibinfo {author} {\bibfnamefont
  {E.}~\bibnamefont {Rotenberg}}, \bibinfo {author} {\bibfnamefont {O.~V.}\
  \bibnamefont {Yazyev}},\ and\ \bibinfo {author} {\bibfnamefont
  {M.}~\bibnamefont {Grioni}},\ }\bibfield  {title} {\bibinfo {title}
  {{Electronic Instability in a Zero-Gap Semiconductor: The Charge-Density Wave
  in $({\mathrm{TaSe}}_{4}{)}_{2}\mathbf{I}$}},\ }\href
  {https://doi.org/10.1103/PhysRevLett.110.236401} {\bibfield  {journal}
  {\bibinfo  {journal} {Phys. Rev. Lett.}\ }\textbf {\bibinfo {volume} {110}},\
  \bibinfo {pages} {236401} (\bibinfo {year} {2013})}\BibitemShut {NoStop}%
\bibitem [{\citenamefont {Gooth}\ \emph {et~al.}(2019)\citenamefont {Gooth},
  \citenamefont {Bradlyn}, \citenamefont {Honnali}, \citenamefont {Schindler},
  \citenamefont {Kumar}, \citenamefont {Noky}, \citenamefont {Qi},
  \citenamefont {Shekhar}, \citenamefont {Sun}, \citenamefont {Wang},
  \citenamefont {Bernevig},\ and\ \citenamefont {Felser}}]{Gooth2019}%
  \BibitemOpen
  \bibfield  {author} {\bibinfo {author} {\bibfnamefont {J.}~\bibnamefont
  {Gooth}}, \bibinfo {author} {\bibfnamefont {B.}~\bibnamefont {Bradlyn}},
  \bibinfo {author} {\bibfnamefont {S.}~\bibnamefont {Honnali}}, \bibinfo
  {author} {\bibfnamefont {C.}~\bibnamefont {Schindler}}, \bibinfo {author}
  {\bibfnamefont {N.}~\bibnamefont {Kumar}}, \bibinfo {author} {\bibfnamefont
  {J.}~\bibnamefont {Noky}}, \bibinfo {author} {\bibfnamefont {Y.}~\bibnamefont
  {Qi}}, \bibinfo {author} {\bibfnamefont {C.}~\bibnamefont {Shekhar}},
  \bibinfo {author} {\bibfnamefont {Y.}~\bibnamefont {Sun}}, \bibinfo {author}
  {\bibfnamefont {Z.}~\bibnamefont {Wang}}, \bibinfo {author} {\bibfnamefont
  {B.~A.}\ \bibnamefont {Bernevig}},\ and\ \bibinfo {author} {\bibfnamefont
  {C.}~\bibnamefont {Felser}},\ }\bibfield  {title} {\bibinfo {title} {{Axionic
  charge-density wave in the Weyl semimetal (TaSe4)2I}},\ }\href
  {https://doi.org/10.1038/s41586-019-1630-4} {\bibfield  {journal} {\bibinfo
  {journal} {Nature}\ }\textbf {\bibinfo {volume} {575}},\ \bibinfo {pages}
  {315} (\bibinfo {year} {2019})}\BibitemShut {NoStop}%
\bibitem [{\citenamefont {Shi}\ \emph {et~al.}(2021)\citenamefont {Shi},
  \citenamefont {Wieder}, \citenamefont {Meyerheim}, \citenamefont {Sun},
  \citenamefont {Zhang}, \citenamefont {Li}, \citenamefont {Shen},
  \citenamefont {Qi}, \citenamefont {Yang}, \citenamefont {Jena}, \citenamefont
  {Werner}, \citenamefont {Koepernik}, \citenamefont {Parkin}, \citenamefont
  {Chen}, \citenamefont {Felser}, \citenamefont {Bernevig},\ and\ \citenamefont
  {Wang}}]{Shi2021}%
  \BibitemOpen
  \bibfield  {author} {\bibinfo {author} {\bibfnamefont {W.}~\bibnamefont
  {Shi}}, \bibinfo {author} {\bibfnamefont {B.~J.}\ \bibnamefont {Wieder}},
  \bibinfo {author} {\bibfnamefont {H.~L.}\ \bibnamefont {Meyerheim}}, \bibinfo
  {author} {\bibfnamefont {Y.}~\bibnamefont {Sun}}, \bibinfo {author}
  {\bibfnamefont {Y.}~\bibnamefont {Zhang}}, \bibinfo {author} {\bibfnamefont
  {Y.}~\bibnamefont {Li}}, \bibinfo {author} {\bibfnamefont {L.}~\bibnamefont
  {Shen}}, \bibinfo {author} {\bibfnamefont {Y.}~\bibnamefont {Qi}}, \bibinfo
  {author} {\bibfnamefont {L.}~\bibnamefont {Yang}}, \bibinfo {author}
  {\bibfnamefont {J.}~\bibnamefont {Jena}}, \bibinfo {author} {\bibfnamefont
  {P.}~\bibnamefont {Werner}}, \bibinfo {author} {\bibfnamefont
  {K.}~\bibnamefont {Koepernik}}, \bibinfo {author} {\bibfnamefont
  {S.}~\bibnamefont {Parkin}}, \bibinfo {author} {\bibfnamefont
  {Y.}~\bibnamefont {Chen}}, \bibinfo {author} {\bibfnamefont {C.}~\bibnamefont
  {Felser}}, \bibinfo {author} {\bibfnamefont {B.~A.}\ \bibnamefont
  {Bernevig}},\ and\ \bibinfo {author} {\bibfnamefont {Z.}~\bibnamefont
  {Wang}},\ }\bibfield  {title} {\bibinfo {title} {{A charge-density-wave
  topological semimetal}},\ }\href {https://doi.org/10.1038/s41567-020-01104-z}
  {\bibfield  {journal} {\bibinfo  {journal} {Nat. Phys.}\ }\textbf {\bibinfo
  {volume} {17}},\ \bibinfo {pages} {381} (\bibinfo {year} {2021})}\BibitemShut
  {NoStop}%
\bibitem [{\citenamefont {Keldysh}\ and\ \citenamefont
  {Kopaev}(1964)}]{Keldysh1964}%
  \BibitemOpen
  \bibfield  {author} {\bibinfo {author} {\bibfnamefont {L.~V.}\ \bibnamefont
  {Keldysh}}\ and\ \bibinfo {author} {\bibfnamefont {Y.~V.}\ \bibnamefont
  {Kopaev}},\ }\bibfield  {title} {\bibinfo {title} {{Possible instability of
  the semimetallic state toward coulomb interaction}},\ }\href@noop {}
  {\bibfield  {journal} {\bibinfo  {journal} {Fiz. Tverd. Tela}\ }\textbf
  {\bibinfo {volume} {6}},\ \bibinfo {pages} {2791} (\bibinfo {year}
  {1964})}\BibitemShut {NoStop}%
\bibitem [{\citenamefont {Peierls}(1955)}]{Peierls1955}%
  \BibitemOpen
  \bibfield  {author} {\bibinfo {author} {\bibfnamefont {R.~E.}\ \bibnamefont
  {Peierls}},\ }\href@noop {} {\emph {\bibinfo {title} {{Quantum Theory of
  Solids}}}}\ (\bibinfo  {publisher} {Oxford University Press},\ \bibinfo
  {address} {Oxford},\ \bibinfo {year} {1955})\BibitemShut {NoStop}%
\bibitem [{\citenamefont {J{\'{e}}rome}\ \emph {et~al.}(1967)\citenamefont
  {J{\'{e}}rome}, \citenamefont {Rice},\ and\ \citenamefont
  {Kohn}}]{Jerome1967}%
  \BibitemOpen
  \bibfield  {author} {\bibinfo {author} {\bibfnamefont {D.}~\bibnamefont
  {J{\'{e}}rome}}, \bibinfo {author} {\bibfnamefont {T.~M.}\ \bibnamefont
  {Rice}},\ and\ \bibinfo {author} {\bibfnamefont {W.}~\bibnamefont {Kohn}},\
  }\bibfield  {title} {\bibinfo {title} {{Excitonic insulator}},\ }\href@noop
  {} {\bibfield  {journal} {\bibinfo  {journal} {Phys. Rev.}\ }\textbf
  {\bibinfo {volume} {158}},\ \bibinfo {pages} {462} (\bibinfo {year}
  {1967})}\BibitemShut {NoStop}%
\bibitem [{\citenamefont {Wei}\ \emph {et~al.}(2012)\citenamefont {Wei},
  \citenamefont {Chao},\ and\ \citenamefont {Aji}}]{Wei2012}%
  \BibitemOpen
  \bibfield  {author} {\bibinfo {author} {\bibfnamefont {H.}~\bibnamefont
  {Wei}}, \bibinfo {author} {\bibfnamefont {S.-P.~P.}\ \bibnamefont {Chao}},\
  and\ \bibinfo {author} {\bibfnamefont {V.}~\bibnamefont {Aji}},\ }\bibfield
  {title} {\bibinfo {title} {{Excitonic Phases from Weyl Semimetals}},\ }\href
  {https://doi.org/10.1103/PhysRevLett.109.196403} {\bibfield  {journal}
  {\bibinfo  {journal} {Phys. Rev. Lett.}\ }\textbf {\bibinfo {volume} {109}},\
  \bibinfo {pages} {196403} (\bibinfo {year} {2012})}\BibitemShut {NoStop}%
\bibitem [{\citenamefont {You}\ \emph {et~al.}(2016)\citenamefont {You},
  \citenamefont {Cho},\ and\ \citenamefont {Hughes}}]{You2016}%
  \BibitemOpen
  \bibfield  {author} {\bibinfo {author} {\bibfnamefont {Y.}~\bibnamefont
  {You}}, \bibinfo {author} {\bibfnamefont {G.~Y.}\ \bibnamefont {Cho}},\ and\
  \bibinfo {author} {\bibfnamefont {T.~L.}\ \bibnamefont {Hughes}},\ }\bibfield
   {title} {\bibinfo {title} {{Response properties of axion insulators and Weyl
  semimetals driven by screw dislocations and dynamical axion strings}},\
  }\href {https://doi.org/10.1103/PhysRevB.94.085102} {\bibfield  {journal}
  {\bibinfo  {journal} {Phys. Rev. B}\ }\textbf {\bibinfo {volume} {94}},\
  \bibinfo {pages} {85102} (\bibinfo {year} {2016})}\BibitemShut {NoStop}%
\bibitem [{\citenamefont {Gr{\"{u}}ner}(1988)}]{Gruner1988a}%
  \BibitemOpen
  \bibfield  {author} {\bibinfo {author} {\bibfnamefont {G.}~\bibnamefont
  {Gr{\"{u}}ner}},\ }\bibfield  {title} {\bibinfo {title} {{The dynamics of
  charge-density waves}},\ }\href {https://doi.org/10.1103/RevModPhys.60.1129}
  {\bibfield  {journal} {\bibinfo  {journal} {Rev. Mod. Phys.}\ }\textbf
  {\bibinfo {volume} {60}},\ \bibinfo {pages} {1129} (\bibinfo {year}
  {1988})}\BibitemShut {NoStop}%
\bibitem [{\citenamefont {Wei}\ \emph {et~al.}(2014)\citenamefont {Wei},
  \citenamefont {Chao},\ and\ \citenamefont {Aji}}]{Wei2014}%
  \BibitemOpen
  \bibfield  {author} {\bibinfo {author} {\bibfnamefont {H.}~\bibnamefont
  {Wei}}, \bibinfo {author} {\bibfnamefont {S.-P.}\ \bibnamefont {Chao}},\ and\
  \bibinfo {author} {\bibfnamefont {V.}~\bibnamefont {Aji}},\ }\bibfield
  {title} {\bibinfo {title} {{Long-range interaction induced phases in Weyl
  semimetals}},\ }\href {https://doi.org/10.1103/PhysRevB.89.235109} {\bibfield
   {journal} {\bibinfo  {journal} {Phys. Rev. B}\ }\textbf {\bibinfo {volume}
  {89}},\ \bibinfo {pages} {235109} (\bibinfo {year} {2014})}\BibitemShut
  {NoStop}%
\bibitem [{\citenamefont {Xue}\ and\ \citenamefont {Zhang}(2017)}]{Xue2017}%
  \BibitemOpen
  \bibfield  {author} {\bibinfo {author} {\bibfnamefont {F.}~\bibnamefont
  {Xue}}\ and\ \bibinfo {author} {\bibfnamefont {X.-X.}\ \bibnamefont
  {Zhang}},\ }\bibfield  {title} {\bibinfo {title} {{Instability and
  topological robustness of Weyl semimetals against Coulomb interaction}},\
  }\href {https://doi.org/10.1103/PhysRevB.96.195160} {\bibfield  {journal}
  {\bibinfo  {journal} {Phys. Rev. B}\ }\textbf {\bibinfo {volume} {96}},\
  \bibinfo {pages} {195160} (\bibinfo {year} {2017})}\BibitemShut {NoStop}%
\bibitem [{\citenamefont {Yang}\ \emph {et~al.}(2011)\citenamefont {Yang},
  \citenamefont {Lu},\ and\ \citenamefont {Ran}}]{Yang2011}%
  \BibitemOpen
  \bibfield  {author} {\bibinfo {author} {\bibfnamefont {K.~Y.}\ \bibnamefont
  {Yang}}, \bibinfo {author} {\bibfnamefont {Y.~M.}\ \bibnamefont {Lu}},\ and\
  \bibinfo {author} {\bibfnamefont {Y.}~\bibnamefont {Ran}},\ }\bibfield
  {title} {\bibinfo {title} {{Quantum Hall effects in a Weyl semimetal:
  Possible application in pyrochlore iridates}},\ }\href
  {https://doi.org/10.1103/PhysRevB.84.075129} {\bibfield  {journal} {\bibinfo
  {journal} {Phys. Rev. B}\ }\textbf {\bibinfo {volume} {84}},\ \bibinfo
  {pages} {075129} (\bibinfo {year} {2011})}\BibitemShut {NoStop}%
\bibitem [{\citenamefont {Breitkreiz}\ and\ \citenamefont
  {Brouwer}(2023)}]{Breitkreiz2022}%
  \BibitemOpen
  \bibfield  {author} {\bibinfo {author} {\bibfnamefont {M.}~\bibnamefont
  {Breitkreiz}}\ and\ \bibinfo {author} {\bibfnamefont {P.~W.}\ \bibnamefont
  {Brouwer}},\ }\bibfield  {title} {\bibinfo {title} {{Fermi-arc metals}},\
  }\href {https://doi.org/10.1103/PhysRevLett.130.196602} {\bibfield  {journal}
  {\bibinfo  {journal} {Phys. Rev. Lett}\ }\textbf {\bibinfo {volume} {130}},\
  \bibinfo {pages} {196602} (\bibinfo {year} {2023})}\BibitemShut {NoStop}%
\bibitem [{\citenamefont {{N I Kulikov}}\ and\ \citenamefont {{V V
  Tugushev}}(1984)}]{Kulikov1984}%
  \BibitemOpen
  \bibfield  {author} {\bibinfo {author} {\bibnamefont {{N I Kulikov}}}\ and\
  \bibinfo {author} {\bibnamefont {{V V Tugushev}}},\ }\bibfield  {title}
  {\bibinfo {title} {{Spin-density waves and itinerant antiferromagnetism in
  metals}},\ }\href {https://doi.org/10.1070/PU1984v027n12ABEH004088}
  {\bibfield  {journal} {\bibinfo  {journal} {Sov. Phys. Uspekhi}\ }\textbf
  {\bibinfo {volume} {27}},\ \bibinfo {pages} {954} (\bibinfo {year}
  {1984})}\BibitemShut {NoStop}%
\bibitem [{\citenamefont {Fawcett}(1988)}]{Fawcett1988a}%
  \BibitemOpen
  \bibfield  {author} {\bibinfo {author} {\bibfnamefont {E.}~\bibnamefont
  {Fawcett}},\ }\bibfield  {title} {\bibinfo {title} {{Spin-density-wave
  antiferromagnetism in chromium}},\ }\href
  {https://doi.org/10.1103/RevModPhys.60.209} {\bibfield  {journal} {\bibinfo
  {journal} {Rev. Mod. Phys.}\ }\textbf {\bibinfo {volume} {60}},\ \bibinfo
  {pages} {209} (\bibinfo {year} {1988})}\BibitemShut {NoStop}%
\bibitem [{\citenamefont {Cho}()}]{Cho2011}%
  \BibitemOpen
  \bibfield  {author} {\bibinfo {author} {\bibfnamefont {G.~Y.}\ \bibnamefont
  {Cho}},\ }\href {https://doi.org/10.48550/ARXIV.1110.1939} {\bibinfo {title}
  {{Possible topological phases of bulk magnetically doped Bi2Se3: turning a
  topological band insulator into the Weyl semimetal}}},\ \Eprint
  {https://arxiv.org/abs/1110.1939} {arXiv:1110.1939} \BibitemShut {NoStop}%
\bibitem [{\citenamefont {Liu}\ \emph {et~al.}(2013)\citenamefont {Liu},
  \citenamefont {Ye},\ and\ \citenamefont {Qi}}]{Liu2013b}%
  \BibitemOpen
  \bibfield  {author} {\bibinfo {author} {\bibfnamefont {C.-X.}\ \bibnamefont
  {Liu}}, \bibinfo {author} {\bibfnamefont {P.}~\bibnamefont {Ye}},\ and\
  \bibinfo {author} {\bibfnamefont {X.-L.}\ \bibnamefont {Qi}},\ }\bibfield
  {title} {\bibinfo {title} {{Chiral gauge field and axial anomaly in a Weyl
  semimetal}},\ }\href {https://doi.org/10.1103/PhysRevB.87.235306} {\bibfield
  {journal} {\bibinfo  {journal} {Phys. Rev. B}\ }\textbf {\bibinfo {volume}
  {87}},\ \bibinfo {pages} {235306} (\bibinfo {year} {2013})}\BibitemShut
  {NoStop}%
\bibitem [{\citenamefont {Vazifeh}\ and\ \citenamefont
  {Franz}(2013)}]{Vazifeh2013}%
  \BibitemOpen
  \bibfield  {author} {\bibinfo {author} {\bibfnamefont {M.~M.}\ \bibnamefont
  {Vazifeh}}\ and\ \bibinfo {author} {\bibfnamefont {M.}~\bibnamefont
  {Franz}},\ }\bibfield  {title} {\bibinfo {title} {{Electromagnetic response
  of Weyl semimetals}},\ }\href
  {https://doi.org/10.1103/PhysRevLett.111.027201} {\bibfield  {journal}
  {\bibinfo  {journal} {Phys. Rev. Lett.}\ }\textbf {\bibinfo {volume} {111}},\
  \bibinfo {pages} {027201} (\bibinfo {year} {2013})}\BibitemShut {NoStop}%
\bibitem [{\citenamefont {Chang}\ \emph {et~al.}(2015)\citenamefont {Chang},
  \citenamefont {Zhou}, \citenamefont {Wang}, \citenamefont {Shan},\ and\
  \citenamefont {Xiao}}]{Chang2015a}%
  \BibitemOpen
  \bibfield  {author} {\bibinfo {author} {\bibfnamefont {H.~R.}\ \bibnamefont
  {Chang}}, \bibinfo {author} {\bibfnamefont {J.}~\bibnamefont {Zhou}},
  \bibinfo {author} {\bibfnamefont {S.~X.}\ \bibnamefont {Wang}}, \bibinfo
  {author} {\bibfnamefont {W.~Y.}\ \bibnamefont {Shan}},\ and\ \bibinfo
  {author} {\bibfnamefont {D.}~\bibnamefont {Xiao}},\ }\bibfield  {title}
  {\bibinfo {title} {{RKKY interaction of magnetic impurities in Dirac and Weyl
  semimetals}},\ }\href {https://doi.org/10.1103/PhysRevB.92.241103} {\bibfield
   {journal} {\bibinfo  {journal} {Phys. Rev. B}\ }\textbf {\bibinfo {volume}
  {92}},\ \bibinfo {pages} {241103(R)} (\bibinfo {year} {2015})}\BibitemShut
  {NoStop}%
\bibitem [{\citenamefont {Wang}\ \emph {et~al.}(2017)\citenamefont {Wang},
  \citenamefont {Chang},\ and\ \citenamefont {Zhou}}]{Wang2017a}%
  \BibitemOpen
  \bibfield  {author} {\bibinfo {author} {\bibfnamefont {S.-X.}\ \bibnamefont
  {Wang}}, \bibinfo {author} {\bibfnamefont {H.-R.}\ \bibnamefont {Chang}},\
  and\ \bibinfo {author} {\bibfnamefont {J.}~\bibnamefont {Zhou}},\ }\bibfield
  {title} {\bibinfo {title} {{RKKY interaction in three-dimensional electron
  gases with linear spin-orbit coupling}},\ }\href
  {https://doi.org/10.1103/PhysRevB.96.115204} {\bibfield  {journal} {\bibinfo
  {journal} {Phys. Rev. B}\ }\textbf {\bibinfo {volume} {96}},\ \bibinfo
  {pages} {115204} (\bibinfo {year} {2017})}\BibitemShut {NoStop}%
\bibitem [{\citenamefont {Nikoli{\'{c}}}(2021{\natexlab{a}})}]{Nikolic2021}%
  \BibitemOpen
  \bibfield  {author} {\bibinfo {author} {\bibfnamefont {P.}~\bibnamefont
  {Nikoli{\'{c}}}},\ }\bibfield  {title} {\bibinfo {title} {{Dynamics of local
  magnetic moments induced by itinerant Weyl electrons}},\ }\href
  {https://doi.org/10.1103/PhysRevB.103.155151} {\bibfield  {journal} {\bibinfo
   {journal} {Phys. Rev. B}\ }\textbf {\bibinfo {volume} {103}},\ \bibinfo
  {pages} {155151} (\bibinfo {year} {2021}{\natexlab{a}})}\BibitemShut
  {NoStop}%
\bibitem [{\citenamefont {Nikoli{\'{c}}}(2021{\natexlab{b}})}]{Nikolic2021b}%
  \BibitemOpen
  \bibfield  {author} {\bibinfo {author} {\bibfnamefont {P.}~\bibnamefont
  {Nikoli{\'{c}}}},\ }\bibfield  {title} {\bibinfo {title} {{Universal spin
  wave damping in magnetic Weyl semimetals}},\ }\href
  {https://doi.org/10.1103/PhysRevB.104.024414} {\bibfield  {journal} {\bibinfo
   {journal} {Phys. Rev. B}\ }\textbf {\bibinfo {volume} {104}},\ \bibinfo
  {pages} {24414} (\bibinfo {year} {2021}{\natexlab{b}})}\BibitemShut {NoStop}%
\bibitem [{\citenamefont {Gaudet}\ \emph {et~al.}(2021)\citenamefont {Gaudet},
  \citenamefont {Yang}, \citenamefont {Baidya}, \citenamefont {Lu},
  \citenamefont {Xu}, \citenamefont {Zhao}, \citenamefont {Rodriguez-Rivera},
  \citenamefont {Hoffmann}, \citenamefont {Graf}, \citenamefont {Torchinsky},
  \citenamefont {Nikoli{\'{c}}}, \citenamefont {Vanderbilt}, \citenamefont
  {Tafti},\ and\ \citenamefont {Broholm}}]{Gaudet2021}%
  \BibitemOpen
  \bibfield  {author} {\bibinfo {author} {\bibfnamefont {J.}~\bibnamefont
  {Gaudet}}, \bibinfo {author} {\bibfnamefont {H.~Y.}\ \bibnamefont {Yang}},
  \bibinfo {author} {\bibfnamefont {S.}~\bibnamefont {Baidya}}, \bibinfo
  {author} {\bibfnamefont {B.}~\bibnamefont {Lu}}, \bibinfo {author}
  {\bibfnamefont {G.}~\bibnamefont {Xu}}, \bibinfo {author} {\bibfnamefont
  {Y.}~\bibnamefont {Zhao}}, \bibinfo {author} {\bibfnamefont {J.~A.}\
  \bibnamefont {Rodriguez-Rivera}}, \bibinfo {author} {\bibfnamefont {C.~M.}\
  \bibnamefont {Hoffmann}}, \bibinfo {author} {\bibfnamefont {D.~E.}\
  \bibnamefont {Graf}}, \bibinfo {author} {\bibfnamefont {D.~H.}\ \bibnamefont
  {Torchinsky}}, \bibinfo {author} {\bibfnamefont {P.}~\bibnamefont
  {Nikoli{\'{c}}}}, \bibinfo {author} {\bibfnamefont {D.}~\bibnamefont
  {Vanderbilt}}, \bibinfo {author} {\bibfnamefont {F.}~\bibnamefont {Tafti}},\
  and\ \bibinfo {author} {\bibfnamefont {C.~L.}\ \bibnamefont {Broholm}},\
  }\bibfield  {title} {\bibinfo {title} {{Weyl-mediated helical magnetism in
  NdAlSi}},\ }\href {https://doi.org/10.1038/s41563-021-01062-8} {\bibfield
  {journal} {\bibinfo  {journal} {Nat. Mater.}\ }\textbf {\bibinfo {volume}
  {20}},\ \bibinfo {pages} {1650} (\bibinfo {year} {2021})}\BibitemShut
  {NoStop}%
\bibitem [{\citenamefont {Yao}\ \emph {et~al.}(2023)\citenamefont {Yao},
  \citenamefont {Gaudet}, \citenamefont {Verma}, \citenamefont {Graf},
  \citenamefont {Yang}, \citenamefont {Bahrami}, \citenamefont {Zhang},
  \citenamefont {Aczel}, \citenamefont {Subedi}, \citenamefont {Torchinsky},
  \citenamefont {Sun}, \citenamefont {Bansil}, \citenamefont {Huang},
  \citenamefont {Singh}, \citenamefont {Nikolic}, \citenamefont {Blaha},\ and\
  \citenamefont {Tafti}}]{Yao2022}%
  \BibitemOpen
  \bibfield  {author} {\bibinfo {author} {\bibfnamefont {X.}~\bibnamefont
  {Yao}}, \bibinfo {author} {\bibfnamefont {J.}~\bibnamefont {Gaudet}},
  \bibinfo {author} {\bibfnamefont {R.}~\bibnamefont {Verma}}, \bibinfo
  {author} {\bibfnamefont {D.~E.}\ \bibnamefont {Graf}}, \bibinfo {author}
  {\bibfnamefont {H.-Y.}\ \bibnamefont {Yang}}, \bibinfo {author}
  {\bibfnamefont {F.}~\bibnamefont {Bahrami}}, \bibinfo {author} {\bibfnamefont
  {R.}~\bibnamefont {Zhang}}, \bibinfo {author} {\bibfnamefont {A.~A.}\
  \bibnamefont {Aczel}}, \bibinfo {author} {\bibfnamefont {S.}~\bibnamefont
  {Subedi}}, \bibinfo {author} {\bibfnamefont {D.~H.}\ \bibnamefont
  {Torchinsky}}, \bibinfo {author} {\bibfnamefont {J.}~\bibnamefont {Sun}},
  \bibinfo {author} {\bibfnamefont {A.}~\bibnamefont {Bansil}}, \bibinfo
  {author} {\bibfnamefont {S.-M.}\ \bibnamefont {Huang}}, \bibinfo {author}
  {\bibfnamefont {B.}~\bibnamefont {Singh}}, \bibinfo {author} {\bibfnamefont
  {P.}~\bibnamefont {Nikolic}}, \bibinfo {author} {\bibfnamefont
  {P.}~\bibnamefont {Blaha}},\ and\ \bibinfo {author} {\bibfnamefont
  {F.}~\bibnamefont {Tafti}},\ }\bibfield  {title} {\bibinfo {title}
  {{Topological spiral magnetism in the Weyl semimetal SmAlSi}},\ }\href
  {https://doi.org/https://doi.org/10.1103/PhysRevX.13.011035} {\bibfield
  {journal} {\bibinfo  {journal} {Phys. Rev. X}\ }\textbf {\bibinfo {volume}
  {13}},\ \bibinfo {pages} {011035} (\bibinfo {year} {2023})}\BibitemShut
  {NoStop}%
\bibitem [{\citenamefont {Drucker}\ \emph {et~al.}(2023)\citenamefont
  {Drucker}, \citenamefont {Nguyen}, \citenamefont {Han}, \citenamefont
  {Siriviboon}, \citenamefont {Luo}, \citenamefont {Andrejevic}, \citenamefont
  {Zhu}, \citenamefont {Bednik}, \citenamefont {Nguyen}, \citenamefont {Chen},
  \citenamefont {Nguyen}, \citenamefont {Liu}, \citenamefont {Williams},
  \citenamefont {Stone}, \citenamefont {Kolesnikov}, \citenamefont {Chi},
  \citenamefont {Fernandez-Baca}, \citenamefont {Nelson}, \citenamefont
  {Alatas}, \citenamefont {Hogan}, \citenamefont {Puretzky}, \citenamefont
  {Huang}, \citenamefont {Yu},\ and\ \citenamefont {Li}}]{Drucker2023}%
  \BibitemOpen
  \bibfield  {author} {\bibinfo {author} {\bibfnamefont {N.~C.}\ \bibnamefont
  {Drucker}}, \bibinfo {author} {\bibfnamefont {T.}~\bibnamefont {Nguyen}},
  \bibinfo {author} {\bibfnamefont {F.}~\bibnamefont {Han}}, \bibinfo {author}
  {\bibfnamefont {P.}~\bibnamefont {Siriviboon}}, \bibinfo {author}
  {\bibfnamefont {X.}~\bibnamefont {Luo}}, \bibinfo {author} {\bibfnamefont
  {N.}~\bibnamefont {Andrejevic}}, \bibinfo {author} {\bibfnamefont
  {Z.}~\bibnamefont {Zhu}}, \bibinfo {author} {\bibfnamefont {G.}~\bibnamefont
  {Bednik}}, \bibinfo {author} {\bibfnamefont {Q.~T.}\ \bibnamefont {Nguyen}},
  \bibinfo {author} {\bibfnamefont {Z.}~\bibnamefont {Chen}}, \bibinfo {author}
  {\bibfnamefont {L.~K.}\ \bibnamefont {Nguyen}}, \bibinfo {author}
  {\bibfnamefont {T.}~\bibnamefont {Liu}}, \bibinfo {author} {\bibfnamefont
  {T.~J.}\ \bibnamefont {Williams}}, \bibinfo {author} {\bibfnamefont {M.~B.}\
  \bibnamefont {Stone}}, \bibinfo {author} {\bibfnamefont {A.~I.}\ \bibnamefont
  {Kolesnikov}}, \bibinfo {author} {\bibfnamefont {S.}~\bibnamefont {Chi}},
  \bibinfo {author} {\bibfnamefont {J.}~\bibnamefont {Fernandez-Baca}},
  \bibinfo {author} {\bibfnamefont {C.~S.}\ \bibnamefont {Nelson}}, \bibinfo
  {author} {\bibfnamefont {A.}~\bibnamefont {Alatas}}, \bibinfo {author}
  {\bibfnamefont {T.}~\bibnamefont {Hogan}}, \bibinfo {author} {\bibfnamefont
  {A.~A.}\ \bibnamefont {Puretzky}}, \bibinfo {author} {\bibfnamefont
  {S.}~\bibnamefont {Huang}}, \bibinfo {author} {\bibfnamefont
  {Y.}~\bibnamefont {Yu}},\ and\ \bibinfo {author} {\bibfnamefont
  {M.}~\bibnamefont {Li}},\ }\bibfield  {title} {\bibinfo {title} {{Topology
  stabilized fluctuations in a magnetic nodal semimetal}},\ }\href
  {https://doi.org/10.1038/s41467-023-40765-1} {\bibfield  {journal} {\bibinfo
  {journal} {Nat. Commun.}\ }\textbf {\bibinfo {volume} {14}},\ \bibinfo
  {pages} {5182} (\bibinfo {year} {2023})}\BibitemShut {NoStop}%
\bibitem [{\citenamefont {Bardeen}\ \emph {et~al.}(1957)\citenamefont
  {Bardeen}, \citenamefont {Cooper},\ and\ \citenamefont
  {Schrieffer}}]{Bardeen1957}%
  \BibitemOpen
  \bibfield  {author} {\bibinfo {author} {\bibfnamefont {J.}~\bibnamefont
  {Bardeen}}, \bibinfo {author} {\bibfnamefont {L.~N.}\ \bibnamefont
  {Cooper}},\ and\ \bibinfo {author} {\bibfnamefont {J.~R.}\ \bibnamefont
  {Schrieffer}},\ }\bibfield  {title} {\bibinfo {title} {{Theory of
  Superconductivity}},\ }\href {https://doi.org/10.1103/PhysRev.108.1175}
  {\bibfield  {journal} {\bibinfo  {journal} {Phys. Rev.}\ }\textbf {\bibinfo
  {volume} {108}},\ \bibinfo {pages} {1175} (\bibinfo {year}
  {1957})}\BibitemShut {NoStop}%
\bibitem [{\citenamefont {Fujikawa}(1979)}]{Fujikawa1979}%
  \BibitemOpen
  \bibfield  {author} {\bibinfo {author} {\bibfnamefont {K.}~\bibnamefont
  {Fujikawa}},\ }\bibfield  {title} {\bibinfo {title} {{Path-Integral Measure
  for Gauge-Invariant Fermion Theories}},\ }\href
  {https://doi.org/10.1103/PhysRevLett.42.1195} {\bibfield  {journal} {\bibinfo
   {journal} {Phys. Rev. Lett.}\ }\textbf {\bibinfo {volume} {42}},\ \bibinfo
  {pages} {1195} (\bibinfo {year} {1979})}\BibitemShut {NoStop}%
\bibitem [{\citenamefont {Schwartz}(2014)}]{Schwartz2014}%
  \BibitemOpen
  \bibfield  {author} {\bibinfo {author} {\bibfnamefont {M.~D.}\ \bibnamefont
  {Schwartz}},\ }\href@noop {} {\emph {\bibinfo {title} {{Quantum field theory
  and the standard model}}}}\ (\bibinfo  {publisher} {Cambridge university
  press},\ \bibinfo {year} {2014})\BibitemShut {NoStop}%
\bibitem [{\citenamefont {Weststr{\"{o}}m}\ and\ \citenamefont
  {Ojanen}(2017)}]{Weststrom2017}%
  \BibitemOpen
  \bibfield  {author} {\bibinfo {author} {\bibfnamefont {A.}~\bibnamefont
  {Weststr{\"{o}}m}}\ and\ \bibinfo {author} {\bibfnamefont {T.}~\bibnamefont
  {Ojanen}},\ }\bibfield  {title} {\bibinfo {title} {{Designer Curved-Space
  Geometry for Relativistic Fermions in Weyl Metamaterials}},\ }\href
  {https://doi.org/10.1103/PhysRevX.7.041026} {\bibfield  {journal} {\bibinfo
  {journal} {Phys. Rev. X}\ }\textbf {\bibinfo {volume} {7}},\ \bibinfo {pages}
  {41026} (\bibinfo {year} {2017})}\BibitemShut {NoStop}%
\bibitem [{\citenamefont {Ilan}\ \emph {et~al.}(2020)\citenamefont {Ilan},
  \citenamefont {Grushin},\ and\ \citenamefont {Pikulin}}]{Ilan2020}%
  \BibitemOpen
  \bibfield  {author} {\bibinfo {author} {\bibfnamefont {R.}~\bibnamefont
  {Ilan}}, \bibinfo {author} {\bibfnamefont {A.~G.}\ \bibnamefont {Grushin}},\
  and\ \bibinfo {author} {\bibfnamefont {D.~I.}\ \bibnamefont {Pikulin}},\
  }\bibfield  {title} {\bibinfo {title} {{Pseudo-electromagnetic fields in 3D
  topological semimetals}},\ }\href {https://doi.org/10.1038/s42254-019-0121-8}
  {\bibfield  {journal} {\bibinfo  {journal} {Nat. Rev. Phys.}\ }\textbf
  {\bibinfo {volume} {2}},\ \bibinfo {pages} {29} (\bibinfo {year}
  {2020})}\BibitemShut {NoStop}%
\bibitem [{\citenamefont {Araki}(2020)}]{Araki2020}%
  \BibitemOpen
  \bibfield  {author} {\bibinfo {author} {\bibfnamefont {Y.}~\bibnamefont
  {Araki}},\ }\bibfield  {title} {\bibinfo {title} {{Magnetic Textures and
  Dynamics in Magnetic Weyl Semimetals}},\ }\href
  {https://doi.org/10.1002/andp.201900287} {\bibfield  {journal} {\bibinfo
  {journal} {Ann. Phys.}\ }\textbf {\bibinfo {volume} {532}},\ \bibinfo {pages}
  {1900287} (\bibinfo {year} {2020})}\BibitemShut {NoStop}%
\bibitem [{\citenamefont {Duan}\ \emph {et~al.}(2018)\citenamefont {Duan},
  \citenamefont {Zheng}, \citenamefont {Fu}, \citenamefont {Wang},
  \citenamefont {Liu}, \citenamefont {Wang},\ and\ \citenamefont
  {Yang}}]{Duan2018}%
  \BibitemOpen
  \bibfield  {author} {\bibinfo {author} {\bibfnamefont {H.-J.}\ \bibnamefont
  {Duan}}, \bibinfo {author} {\bibfnamefont {S.-H.}\ \bibnamefont {Zheng}},
  \bibinfo {author} {\bibfnamefont {P.-H.}\ \bibnamefont {Fu}}, \bibinfo
  {author} {\bibfnamefont {R.-Q.}\ \bibnamefont {Wang}}, \bibinfo {author}
  {\bibfnamefont {J.-F.}\ \bibnamefont {Liu}}, \bibinfo {author} {\bibfnamefont
  {G.-H.}\ \bibnamefont {Wang}},\ and\ \bibinfo {author} {\bibfnamefont
  {M.}~\bibnamefont {Yang}},\ }\bibfield  {title} {\bibinfo {title} {{Indirect
  magnetic interaction mediated by Fermi arc and boundary reflection near Weyl
  semimetal surface}},\ }\href@noop {} {\bibfield  {journal} {\bibinfo
  {journal} {New J. Phys.}\ }\textbf {\bibinfo {volume} {20}},\ \bibinfo
  {pages} {103008} (\bibinfo {year} {2018})}\BibitemShut {NoStop}%
\bibitem [{\citenamefont {Kaladzhyan}\ \emph {et~al.}(2019)\citenamefont
  {Kaladzhyan}, \citenamefont {Zyuzin},\ and\ \citenamefont
  {Simon}}]{Kaladzhyan2019a}%
  \BibitemOpen
  \bibfield  {author} {\bibinfo {author} {\bibfnamefont {V.}~\bibnamefont
  {Kaladzhyan}}, \bibinfo {author} {\bibfnamefont {A.~A.}\ \bibnamefont
  {Zyuzin}},\ and\ \bibinfo {author} {\bibfnamefont {P.}~\bibnamefont
  {Simon}},\ }\bibfield  {title} {\bibinfo {title} {{RKKY interaction on the
  surface of three-dimensional Dirac semimetals}},\ }\href
  {https://doi.org/10.1103/PhysRevB.99.165302} {\bibfield  {journal} {\bibinfo
  {journal} {Phys. Rev. B}\ }\textbf {\bibinfo {volume} {99}},\ \bibinfo
  {pages} {165302} (\bibinfo {year} {2019})}\BibitemShut {NoStop}%
\bibitem [{\citenamefont {Verma}\ \emph {et~al.}(2020)\citenamefont {Verma},
  \citenamefont {Giri}, \citenamefont {Fertig},\ and\ \citenamefont
  {Kundu}}]{Verma2020}%
  \BibitemOpen
  \bibfield  {author} {\bibinfo {author} {\bibfnamefont {S.}~\bibnamefont
  {Verma}}, \bibinfo {author} {\bibfnamefont {D.}~\bibnamefont {Giri}},
  \bibinfo {author} {\bibfnamefont {H.~A.}\ \bibnamefont {Fertig}},\ and\
  \bibinfo {author} {\bibfnamefont {A.}~\bibnamefont {Kundu}},\ }\bibfield
  {title} {\bibinfo {title} {{RKKY coupling in Weyl semimetal thin films}},\
  }\href {https://doi.org/10.1103/PhysRevB.101.085419} {\bibfield  {journal}
  {\bibinfo  {journal} {Phys. Rev. B}\ }\textbf {\bibinfo {volume} {101}},\
  \bibinfo {pages} {85419} (\bibinfo {year} {2020})}\BibitemShut {NoStop}%
\bibitem [{\citenamefont {Li}\ \emph {et~al.}(2019)\citenamefont {Li},
  \citenamefont {Li}, \citenamefont {Du}, \citenamefont {Wang}, \citenamefont
  {Gu}, \citenamefont {Zhang}, \citenamefont {He}, \citenamefont {Duan},\ and\
  \citenamefont {Xu}}]{A02Li2019}%
  \BibitemOpen
  \bibfield  {author} {\bibinfo {author} {\bibfnamefont {J.}~\bibnamefont
  {Li}}, \bibinfo {author} {\bibfnamefont {Y.}~\bibnamefont {Li}}, \bibinfo
  {author} {\bibfnamefont {S.}~\bibnamefont {Du}}, \bibinfo {author}
  {\bibfnamefont {Z.}~\bibnamefont {Wang}}, \bibinfo {author} {\bibfnamefont
  {B.-L.}\ \bibnamefont {Gu}}, \bibinfo {author} {\bibfnamefont {S.-C.}\
  \bibnamefont {Zhang}}, \bibinfo {author} {\bibfnamefont {K.}~\bibnamefont
  {He}}, \bibinfo {author} {\bibfnamefont {W.}~\bibnamefont {Duan}},\ and\
  \bibinfo {author} {\bibfnamefont {Y.}~\bibnamefont {Xu}},\ }\bibfield
  {title} {\bibinfo {title} {{Intrinsic magnetic topological insulators in van
  der Waals layered MnBi2Te4-family materials}},\ }\href
  {https://doi.org/10.1126/sciadv.aaw5685} {\bibfield  {journal} {\bibinfo
  {journal} {Sci. Adv.}\ }\textbf {\bibinfo {volume} {5}},\ \bibinfo {pages}
  {eaaw5685} (\bibinfo {year} {2019})}\BibitemShut {NoStop}%
\bibitem [{\citenamefont {Otrokov}\ \emph {et~al.}(2019)\citenamefont
  {Otrokov}, \citenamefont {Klimovskikh}, \citenamefont {Bentmann},
  \citenamefont {Estyunin}, \citenamefont {Zeugner}, \citenamefont {Aliev},
  \citenamefont {Ga{\ss}}, \citenamefont {Wolter}, \citenamefont {Koroleva},
  \citenamefont {Shikin},\ and\ \citenamefont {Others}}]{Otrokov2019MBT}%
  \BibitemOpen
  \bibfield  {author} {\bibinfo {author} {\bibfnamefont {M.~M.}\ \bibnamefont
  {Otrokov}}, \bibinfo {author} {\bibfnamefont {I.~I.}\ \bibnamefont
  {Klimovskikh}}, \bibinfo {author} {\bibfnamefont {H.}~\bibnamefont
  {Bentmann}}, \bibinfo {author} {\bibfnamefont {D.}~\bibnamefont {Estyunin}},
  \bibinfo {author} {\bibfnamefont {A.}~\bibnamefont {Zeugner}}, \bibinfo
  {author} {\bibfnamefont {Z.~S.}\ \bibnamefont {Aliev}}, \bibinfo {author}
  {\bibfnamefont {S.}~\bibnamefont {Ga{\ss}}}, \bibinfo {author} {\bibfnamefont
  {A.~U.~B.}\ \bibnamefont {Wolter}}, \bibinfo {author} {\bibfnamefont {A.~V.}\
  \bibnamefont {Koroleva}}, \bibinfo {author} {\bibfnamefont {A.~M.}\
  \bibnamefont {Shikin}},\ and\ \bibinfo {author} {\bibnamefont {Others}},\
  }\bibfield  {title} {\bibinfo {title} {{Prediction and observation of an
  antiferromagnetic topological insulator}},\ }\href
  {https://doi.org/10.1038/s41586-019-1840-9} {\bibfield  {journal} {\bibinfo
  {journal} {Nature}\ }\textbf {\bibinfo {volume} {576}},\ \bibinfo {pages}
  {416} (\bibinfo {year} {2019})}\BibitemShut {NoStop}%
\bibitem [{\citenamefont {Deng}\ \emph {et~al.}(2017)\citenamefont {Deng},
  \citenamefont {Luo}, \citenamefont {Wang}, \citenamefont {Sheng},\ and\
  \citenamefont {Xing}}]{Deng2017}%
  \BibitemOpen
  \bibfield  {author} {\bibinfo {author} {\bibfnamefont {M.~X.}\ \bibnamefont
  {Deng}}, \bibinfo {author} {\bibfnamefont {W.}~\bibnamefont {Luo}}, \bibinfo
  {author} {\bibfnamefont {R.~Q.}\ \bibnamefont {Wang}}, \bibinfo {author}
  {\bibfnamefont {L.}~\bibnamefont {Sheng}},\ and\ \bibinfo {author}
  {\bibfnamefont {D.~Y.}\ \bibnamefont {Xing}},\ }\bibfield  {title} {\bibinfo
  {title} {{Weyl semimetal induced from a Dirac semimetal by magnetic
  doping}},\ }\href {https://doi.org/10.1103/PhysRevB.96.155141} {\bibfield
  {journal} {\bibinfo  {journal} {Phys. Rev. B}\ }\textbf {\bibinfo {volume}
  {96}},\ \bibinfo {pages} {155141} (\bibinfo {year} {2017})}\BibitemShut
  {NoStop}%
\bibitem [{\citenamefont {Cano}\ \emph {et~al.}(2017)\citenamefont {Cano},
  \citenamefont {Bradlyn}, \citenamefont {Wang}, \citenamefont {Hirschberger},
  \citenamefont {Ong},\ and\ \citenamefont {Bernevig}}]{Cano2017}%
  \BibitemOpen
  \bibfield  {author} {\bibinfo {author} {\bibfnamefont {J.}~\bibnamefont
  {Cano}}, \bibinfo {author} {\bibfnamefont {B.}~\bibnamefont {Bradlyn}},
  \bibinfo {author} {\bibfnamefont {Z.}~\bibnamefont {Wang}}, \bibinfo {author}
  {\bibfnamefont {M.}~\bibnamefont {Hirschberger}}, \bibinfo {author}
  {\bibfnamefont {N.~P.}\ \bibnamefont {Ong}},\ and\ \bibinfo {author}
  {\bibfnamefont {B.~A.}\ \bibnamefont {Bernevig}},\ }\bibfield  {title}
  {\bibinfo {title} {{Chiral anomaly factory: Creating Weyl fermions with a
  magnetic field}},\ }\href {https://doi.org/10.1103/PhysRevB.95.161306}
  {\bibfield  {journal} {\bibinfo  {journal} {Phys. Rev. B}\ }\textbf {\bibinfo
  {volume} {95}},\ \bibinfo {pages} {161306(R)} (\bibinfo {year}
  {2017})}\BibitemShut {NoStop}%
\end{thebibliography}%

\onecolumngrid

\appendix

\section{Density operator and Polarization}

The wavefunctions of the Fermi-arc metal are given in Eq.\ \eqref{wf} of the main text as
\begin{gather}
\psi_{\bm{k}\chi} = \bigg(\frac{\pi K}{4Q}\bigg)^{\frac{1}{4}} \begin{pmatrix}
  1\\ - e^{i\phi_{\bm{k}}}  \end{pmatrix}
\sum_n e^{i(\bm{k}_\parallel \cdot\bm{r}_\parallel +k_z z_{\bm{k}n\chi})} e^{-\frac{KQ}{2}(z - z_{\bm{k},n,\chi})^2} \ \ , \nonumber\\
z_{\bm{k}n+} =
\frac{\phi_{\bm{k}} + 2\pi n}{Q} \ \ ,  \ \  z_{\bm{k}n-} = z_{\bm{k}n+} + \frac{\pi}{Q}.
\label{swf}
\end{gather}
For the density operator follows
\begin{gather}
\rho(\vq) = \sum_{\vk\vk'\chi} c^\dagger_{\vk\chi}c^\pd_{\vk'\chi}   \int d\rr  \ e^{-i\vq\cdot\rr}\ 
  \Psi^\dagger_{\vk'\chi}(\rr)\Psi^\pd_{\vk\chi}(\rr)\\
  = \sum_{\vk \chi }c^\dagger_{\vk\chi}c^\pd_{\vk+\vq,\chi}\  e^{-\frac{K}{4Q}(\phi_{\vk}-\phi_{\vk+\vq})^2},
\end{gather}
where we approximated (used also below)   
\begin{equation}
    \frac{1+e^{i(\phi_{\vk}-\phi_{\vk+\vq})}}{2} e^{-\frac{K}{4Q}(\phi_{\vk}-\phi_{\vk+\vq})^2} \approx
     e^{-\frac{K}{4Q}(\phi_{\vk}-\phi_{\vk+\vq})^2}
\end{equation}
due to $Q\ll K$. The physical reason for the unusual form factor $\exp\{-\frac{K}{4Q}(\phi_{\vk}-\phi_{\vk+\vq})^2\}$ in the density is that two wavefunctions at different momenta (more precisely at different polar angles $\phi_\vk$) are localized at different $z$ slices so that they don't interfere and thus don't produce a density wave. 

The static polarization is given by
\begin{gather}
\Pi(\vq)\equiv\Pi(\bm{q},\Omega=0)  = -i\int_0^\infty dt \  \Big\langle \big[ \rho(\vq,t),\rho(-\vq,0)
\big]\Big\rangle \\
= \ -i\int_{0}^\infty dt \   \sum_{\vk\vk'\chi\chi'}  
\Big\langle \big[ c^\dagger_{\vk\chi}c^\pd_{\vk+\vq\chi},c^\dagger_{\vk'\chi'}c^\pd_{\vk'-\vq\chi ' }
\big]\Big\rangle \nonumber \\\ \ \ \ \ \  \times
e^{i(\en_{\vk\chi}-\en_{\vk+\vq,\chi})t}
e^{-\frac{K}{4Q}(\phi_\vk-\phi_{\vk+\vq})^2}
e^{-\frac{K}{4Q}(\phi_{\vk'}-\phi_{\vk'-\vq})^2}
\\
=  \sum_{\bm{k},\chi}
\frac{n_F(\varepsilon_{\bm{k},\chi}) - n_F(\varepsilon_{\bm{k}+\bm{q},\chi})}{\varepsilon_{\bm{k},\chi}- \varepsilon_{\bm{k}+\bm{q},\chi}}e^{-\frac{K}{2Q}(\phi_\vk-\phi_{\vk+\vq})^2},
\end{gather}
where $n_F(x)=1/(1+\exp{(x/T)})$ is the Fermi-Dirac distribution and the Boltzmann constant set to unity. 

The density-caused form factor $\exp\{-\frac{K}{2Q}(\phi_\vk-\phi_{\vk+\vq})^2\}$ describes the suppressed interference of particles separated in $z$ and thus a suppressed ability of the system to polarize at momenta $|\vq_\parallel|\gtrsim\sqrt{KQ}$.
In the limit $|\vq_\parallel| \ll \sqrt{KQ}$, we instead recover the usual Thomas-Fermi limit
\begin{gather}
\Pi(\bm{q})\rightarrow -\nu = -\frac{KQ}{2\pi^2 v} \ \ , \label{stf}
\end{gather}
where $\nu=KQ/2\pi^2v$ is the density of states at the Fermi energy.

\section{Mean field decoupling}

The density-density interaction potential taken in the static limit and with a RPA-screened interaction vertex, as motivated in the main text, reads
\begin{gather}
    V=\frac{1}{2}\sum_{\vq} \frac{U_{\vq}}{1-\Pi(\vq)U_{\vq}} \rho(\vq)\rho(-\vq).
\end{gather}
Mean-field decoupling in the excitonic channel corresponds to the coupling of opposite chiralities, $\chi=+$ and $\chi=-$, as those have electron-like and hole-like dispersions, respectively, $\en_{\vk\chi}\equiv \chi \, v (|\vk_\parallel| - K) $. The mean-field Hamiltonian then reads
\begin{gather}
      H_{\mathrm{MF}} = \sum_{\vk\chi} \en_{\vk\chi} 
    c^\dagger_{\vk\chi}c^\pd_{\vk\chi} +
    \sum_\vk c^\dagger_{\vk,+}c^\pd_{\vk,-}
    \Delta^\ast_{+-}(\vk)+\sum_\vk c^\dagger_{\vk,-}c^\pd_{\vk,+}
    \Delta_{+-}(\vk),\label{smf}
\end{gather}
with 
\begin{gather}
    \Delta_{+-}(\vk) = -\sum_{\vk'} \Gamma_{+-}(\vk,\vk')\langle c^\dagger_{\vk',+}c^\pd_{\vk',-}\rangle,
    \label{sdeq}
\end{gather}
and the interaction vertex
\begin{gather}
\Gamma_{+-}(\bm{k},\bm{k}') =\frac{U_{\vk-\vk'}}{1 -\Pi(\vk-\vk')U_{\vk-\vk'}} e^{-\frac{K}{2Q}(\phi_\vk-\phi_{\vk'})^2}.
\label{sgam}
\end{gather}
It follows the standard derivation of the self-consistent gap equation via diagonolizing the mean-field Hamiltonian and inserting into \eqref{sdeq},  
\begin{gather}
    \Delta_{+-}(\vk) = \sum_{\vk'} \Gamma_{+-}(\vk,\vk') \frac{\Delta_{+-}(\vk')}{2 \zeta_\vk}\left[ n_F(\zeta_\vk)-n_F(-\zeta_\vk)\right],\ \ \ \zeta_\vk = \sqrt{\en_{\vk'\chi}^2+\Delta_{+-}^2(\vk')}.
\end{gather}
At $T_c$ the order parameter vanishes, leading to  
\begin{equation}
    1=\sum_{\vk'}\Gamma_{+-}(\vk,\vk')\frac{1}{2|\en_{\vk'\chi}|}\tanh\frac{|\en_{\vk'\chi}|}{2T_c},
    \label{stc}
\end{equation}
to be solved with respect to $T_c$.

\section{Derivation of $T_c$}

The form factor in the interaction vertex modifies the otherwise standard solving of \eqref{stc} with respect to $T_c$, which we are now going to perform. We use that  the interaction vertex, given in \eqref{sgam},
is suppressed for $ |\vq_\parallel|\gtrsim\sqrt{KQ}$ and we may use the Thomas-Fermi limit \eqref{stf} for the polarization, leading to 
\begin{gather}
\Gamma_{+-}(\bm{k},\bm{k}') \rightarrow \frac{U_{\bm{k}-\bm{k}'}}{1 + \nu U_{\bm{k}-\bm{k}'}}e^{-\frac{K}{2Q}(\phi_\vk-\phi_{\vk'})^2}\ \ .
\end{gather}
Inserting $U_{\vq} = 4\pi e^2/q^2 $, $\nu = KQ/2\pi^2 v$, setting $\vk=(K,0,0)$ (without loss of generality), and using polar  coordinates $\vk' = (k'\cos\phi',k'\sin\phi',k_z')$ we obtain
\begin{gather}
1= \frac{e^2}{2\pi^2}\int dk_z' \int d\phi'\int dk'k'\frac{e^{-\frac{K}{2Q}\phi'{}^2} }{k_z'{}^2+(k'-K)^2+k'K\phi'{}^2+2\xi^2 QK}\ \frac{\tanh\frac{v|k'-K|}{2T_c}}{2v|k'-K|}
\end{gather}
where $\xi^2 = e^2/v\pi \approx (0.002 c/v)^2$.
Using that the $\tanh$ term is peaked at $k'=K$ with a width $T_c^2 \ll QK$, and $k_z^2\sim Q^2 \ll QK$, the integrals factorize and can be carried out, 
\begin{gather}
1= \frac{\xi^2 Q}{2\pi K} \int d\phi'\frac{e^{-\frac{K}{2Q}\phi'{}^2} }{\phi'{}^2+\xi^2 \frac{2Q}{K}}\ \int d\en \frac{\tanh\frac{|\en|}{2T_c}}{2|\en|}\nonumber \\
\ \ \ \ \ \  =\eta \sqrt{\frac{Q}{K}}\left(\ln\frac{w}{2T_c}+\gamma-\ln\frac{\pi}{4}\right),\ \ \ \eta=\frac{\xi\, e^{\xi^2}\mathrm{erf}\,\xi}{ 2\sqrt{2}},\\
\Longrightarrow T_c =  \frac{2e^\gamma}{\pi} w \exp \left[-\frac{1}{\eta}\sqrt{\frac{K}{Q}}\right] 
\end{gather}
where $w$ is the UV energy cut-off and $\gamma$ the Euler constant.

\section{Derivation of Landau-Ginzburg theory for excitons}

Expanding the free energy to quadratic order a complex order parameter $\Delta$ appearing in the mean-field Hamiltonian \eqref{smf} we find
\begin{gather}
F =\underbrace{ \left[\frac{1}{g} + \Pi\right]}_{=\alpha}|\Delta|^2 + \frac{\beta}{2} |\Delta|^4 
\label{seq1}
\end{gather}
where $g$ is an effective coupling constant which may be related to the $T_c$ derived in the main text, and
\begin{gather}
\Pi = T \sum_\omega \sum_{\bm{k}} G_+(\omega,\bm{k}) G_-(\omega,\bm{k}) =- T \sum_\omega \sum_{\bm{k}} \frac{1}{\omega^2 + \varepsilon_{\bm{k},+}^2} \ \ , \nonumber \\
\beta = T \sum_\omega \sum_{\bm{k}} (G_+(\omega,\bm{k}) G_-(\omega,\bm{k}))^2 = T \sum_\omega \sum_{\bm{k}}\frac{1}{(\omega^2 + \varepsilon_{\bm{k},+}^2)^2}
\end{gather}
where
\begin{gather}
G_\pm(\omega,\bm{k}) = \frac{1}{i\omega - \varepsilon_{\bm{k},\pm}} \ \ , \nonumber \\
\varepsilon_{\bm{k},+} =- \varepsilon_{\bm{k},-} = v(k_\parallel - b).
\end{gather}

For the first sum, we first perform the Matsubara summation using the relation $T\sum_\omega g(i\omega) = \oint n_F(\omega)g(\omega) \frac{d\omega}{2\pi i}$ where the contour encloses the entire complex plane anticlockwise, excluding the imaginary axis. Taking into account the residues at $i\omega = \pm \varepsilon_{\bm{k},+}$ we find
\begin{gather}
\Pi = \sum_{\bm{k}} \frac{n_F(\varepsilon_{\bm{k},+}) - n_F(\varepsilon_{\bm{k},-})}{\varepsilon_{\bm{k},+} - \varepsilon_{\bm{k},-}} = \int \frac{n_F(\varepsilon) - n_F(-\varepsilon)}{2\varepsilon} \nu(\varepsilon)d\varepsilon \nonumber \\
\approx- \nu\int_{-\Lambda}^{\Lambda} \frac{\tanh(\frac{\varepsilon}{2T})}{2\varepsilon} d\varepsilon \approx -\nu \ln(\frac{\Lambda}{T}) \ \ ,
\end{gather}
where $\nu(\varepsilon)$ is the density of states and $\nu=\nu(0)$ is the density of states at the Fermi energy. Inserting into \eqref{seq1} we obtain
\begin{gather}
\alpha = \frac{1}{g} -\nu \ln(\frac{\Lambda}{T}) = -\nu \ln(\frac{T_c}{T}) \approx \nu \left[1-\frac{T_c}{T}\right].
\end{gather}

For the second sum we first perform the integral over momentum and obtain
\begin{gather}
\beta = T \sum_\omega \int \frac{\nu(\varepsilon)d \varepsilon}{(\omega^2+\varepsilon^2)^2} \approx \frac{\pi \nu T}{2}\sum_\omega \frac{1}{|\omega|^3} = \frac{7\zeta(3)}{8\pi^2}\frac{\nu}{T^2},
\end{gather}
where we used
\begin{gather}
\sum_{\omega = 2\pi(n+\frac{1}{2})T} \frac{1}{|\omega|^3} = \frac{2}{\pi^3} \frac{7\zeta(3)}{8T^3} \ \ , \ \ \int \frac{d\varepsilon}{(\omega^2+\varepsilon^2)^2} = \frac{\pi}{|\omega|^3}.
\end{gather}

In the excitonic phase, i.e., below $T_c$, the order parameter becomes $|\Delta|^2 =-\alpha/\beta $  and the Fermi energy thus assumes the value
\begin{gather}
    F_\Delta = -\frac{\alpha^2}{2\beta} = -\nu\big(T-T_c\big)^2\frac{4\pi^2}{7\zeta(3)}.
\end{gather}

\end{document}